\documentclass[11pt]{iopart}
\usepackage{iopams}
\expandafter\let\csname equation*\endcsname\relax
\expandafter\let\csname endequation*\endcsname\relax
\usepackage{amsmath,amssymb,amsfonts,amsthm}
\usepackage{amssymb}
\usepackage{mathtools}
\usepackage{cancel} 
\usepackage{bm}
\usepackage[usenames, dvipsnames]{color} 
\usepackage{dcolumn}
\usepackage{epic} 
\usepackage{epsfig}
\usepackage{feynmp}
\usepackage{graphicx}
\usepackage{grffile}
\usepackage[breaklinks,colorlinks = true,linkcolor = red,urlcolor=blue,citecolor=cyan]{hyperref}
\usepackage{caption}
\usepackage{subcaption}
\usepackage{mathrsfs}
\usepackage{soul}
\usepackage{wrapfig}
\usepackage{xy} 
\usepackage{xcolor}
\usepackage[ascii]{inputenc}
\usepackage{makecell}
\usepackage[mathscr]{eucal}
\usepackage{cite}
\usepackage{textcomp}
\usepackage{bbold}
\usepackage{ulem}

\usepackage[mathscr]{eucal}
\usepackage{color}
\usepackage{dsfont}

\definecolor{dgreen}{rgb}{0,0.7,0}

\newcommand{\titlename}{Conserved densities of hard rods: microscopic to hydrodynamic solutions}

\begin{document}
\title{\titlename}
\author{Mrinal Jyoti Powdel}
\address{International Centre for Theoretical Sciences, Tata Institute of Fundamental Research, Bengaluru -- 560089, India}
\author{Anupam Kundu}
\address{International Centre for Theoretical Sciences, Tata Institute of Fundamental Research, Bengaluru -- 560089, India}
\ead{mrinal.jyoti@icts.res.in, anupam.kundu@icts.res.in}

\begin{abstract}
We consider a system of many hard rods moving in {one dimension}. As it is an integrable system, it possesses an extensive number of conserved quantities and  its evolution on macroscopic scale can be described by generalised hydrodynamics. Using a microscopic approach, we compute the evolution of the conserved densities starting from non-equilibrium initial conditions of both quenched and annealed type. In addition to getting reduced to the Euler solutions of  the hydrodynamics in the thermodynamic limit, the microscopic solutions can also capture effects of the Navier-Stokes terms and thus go beyond the Euler solutions. We demonstrate this feature {from microscopic analysis and numerical solution of the Navier-Stokes equation} in two problems --- first, tracer diffusion in a background of hard rods and second, the evolution from a domain wall initial condition in which the velocity distribution of the rods are different on the two sides of the interface. We supplement our analytical results using extensive numerical simulations.

\end{abstract}	

\maketitle

\section{Introduction}
The study of dynamics of interacting systems starting from non-equilibrium initial states is an important problem in statistical physics. Generic classical Hamiltonian systems starting from low entropy initial conditions evolve to thermal Gibbs state characterized by few conserved densities of the dynamics, and their evolution is governed by the hydrodynamic evolution of those conserved densities. An example of such initial conditions is  domain wall initial condition or, more generally,  one with inhomogeneous density profiles on  macroscopic scales. One is generally interested in the hydrodynamic evolution of {the system from} such inhomogeneous initial conditions and its approach to the global equilibrium. While such problems have been well studied in the context of non-integrable systems, there has recently grown a huge interest for investigating such questions in the context of integrable systems. The scenario is different for integrable systems as they {possess} extensive number of conserved quantities.  Such systems are believed to thermalise to Generalised Gibbs Ensemble (GGE) \cite{rigol2007relaxation, pozsgay2013generalized, langen2015experimental, vidmar2016generalized}. Experiments as well as theoretical studies in the last decade have led to the development of the generalised version of thermodynamics and hydrodynamics for integrable systems \cite{kinoshita2006quantum, malvania2021generalized,castro2016emergent,bertini2016transport,doyon2019generalized,doyonlecturenotes, spohn2018interacting, spohn2024hydrodynamic}. The theoretical formulation has provided a fundamental tool to describe and understand the evolution of quantum integrable models in one dimension starting from general inhomogeneous state and thermalisation to generalised Gibbs state \cite{doyonlecturenotes, bastianello2022introduction, alba2021generalized, caux2019hydrodynamics, cassidy2011generalized, calabrese2011quantum, eisert2021entangling}. While there have been a number of studies in quantum context \cite{rigol2007relaxation, alba2021generalized, schemmer2019generalized, ruggiero2020quantum}, such studies in the context of classical integrable systems are relatively small. Among the few classical integrable models for which GHD formulation has been studied, a collection of hard rods in one dimension is a simple yet non trivial interacting integrable system that displays most of the essential features both in equilibrium and non-equilibrium states.

A {hard rod} system consists of $N$ identical hard rods each of unit mass and length $a$ moving on a {one-dimensional} line which could be of finite or infinite length. The positions and {velocities} of the rods are denoted by $\{X_i,V_i\}$ for $i=1,2,...,N$. Since mass of each rod is unity, the momentum of each rod is its velocity and hence, {the momenta are also} denoted by $V_i, ~i=1,2,...,N$. The rods move ballistically in between elastic collisions such that when two such rods collide, they exchange their momenta. 
Since the momenta of two rods are only exchanged in the event of a collision, the sum of the powers of their momenta remains conserved. {Generalizing this to the full system on an infinite line, we obtain that the sum of the powers of momenta of all the hard rods denoted by 
$Q_\alpha = \sum_{i=1}^N V_i^\alpha$ with $\alpha=1,2,..,N$ are conserved and satisfy zero Poisson commutation relations \cite{OLSHANETSKY1981313}. One should note that, in fact, any function, $S(V_1, V_2, \cdots V_N)$ which is symmetric with respect to the exchange of velocities is also conserved. However, to obtain a hydrodynamic description, one needs to choose a set of local conserved quantities  \cite{doyonlecturenotes} like $\{ Q_\alpha \}$  mentioned above. }

In 1930, the hard rod system was introduced by L. Tonks (hence, also known as Tonks gas) in the context of  classical equilibrium fluid \cite{tonks1936complete} for which an equation of state, different from that of an ideal gas, was computed analytically. This model and its variants later have been used as examples of simple interacting systems for which other quantities such as density fluctuations, pair correlation functions, distribution functions were computed analytically in thermal equilibrium state \cite{tonks1936complete, robledo1986distribution, salsburg1953molecular, sells1953pair, percus1964the, koppel1963partition}.  Later, time evolution in hard rod systems was studied by deriving hydrodynamic equations from Boltzmann equation \cite{percus1969exact} as well as from microscopic computations \cite{lebowitz1968time, bernstein1988expansion, jepsen1965dynamics, valleau1970time}.

Recently, the {hard rod} system has seen a revival of interest with the development of GHD  which is written for an integrable system in terms of phase space density of the quasiparticles \cite{doyonlecturenotes, spohn2024hydrodynamic}. These are particles  with tracers that jump from one particle to another such that the tracer particle follows a given `asymptotic momentum'. For hard rod system,  quasiparticles are  rods that exchange labels (instead of momenta) at each collision. The GHD for hard rods can be written in terms of the single particle phase space distribution $f(X,V,t)$ defined as 
\begin{align}
f(X,V,t) &= \sum_{i=1}^N \left \langle \delta\left (X-X_i(t)\right ) \delta(V-{V}_i) \right \rangle, \label{def:f(X,V,t)}
\end{align}
where the average $\langle \cdots \rangle$ is taken over an ensemble of initial conditions corresponding to specified macroscopic profiles of conserved densities such as mass density, momentum density and/or other conserved densities.
The GHD equation at the Euler scale is written as 
\begin{equation}\label{eq:GHD}
    \partial_t f(X,V,t) + \partial_X V_{\rm eff}(X,V,t) f(X,V,t) =0 \ .
\end{equation}
{\color{black}
Here, $V_{\rm eff}$ is the effective velocity with which a quasiparticle with bare velocity $V$ moves and is given by
\begin{equation}\label{def:V_eff}
    V_{\rm eff}(X,V,t) = \frac{V-a\rho(X,t) U(X,t)}{(1-a\rho(X,t))} \ , 
\end{equation}
with $\rho(X,t)$ denoting the hard rod mass density,  
\begin{equation}
    \rho(X,t)=\int dV~f(X,V,t) \ ,
\end{equation}
and $U(X,t)$ denoting the velocity of the hard rod flow at position $X$ and time $t$,
\begin{equation}\label{def:rho-U}
    U(X,t) = \frac{\int dV~Vf(X,V,t) }{\rho(X,t)} \ .
\end{equation}
{It should be mentioned here that i}n the context of quantum integrable systems, the Euler scale GHD in the form of Eq.~\eqref{eq:GHD} was derived in \cite{castro2016emergent,bertini2016transport} using thermodynamic Bethe ansatz {and here} $f(X,V,t)$ represents the local density of quasiparticle excitations  occupying pseudomomenta $V$ at time $t$. The Euler {scale} GHD equation has also been viewed  as collisionless Boltzmann equation \cite{bulchandani2018bethe}. Long before the development of GHD as in \cite{castro2016emergent,bertini2016transport},  a similar form of Euler hydrodynamic equation was obtained for classical soliton gas using kinetic theory\cite{zakharov1971kinetic, el2005kinetic}. Since the computation of $V_{\rm eff}$ requires a Bethe solutions of the underlying quantum system, the Euler GHD equation of the form Eq.~\eqref{eq:GHD} has recently been called {as the} Bethe-Boltzmann equation \cite{bulchandani2018bethe}.}

{In hard rod systems, e}very elastic collision makes a {quasiparticle} jump by an distance equal to its rod length. Due to the cumulative effect of these jumps over a time duration $t$, a {quasiparticle} with {bare} velocity $V$ acquires an extra displacement in addition to the displacement, $Vt$ due to drift.  While the total number of jumps gives rise to an effective velocity of the quasiparticle, the fluctuations {in} the number of jumps give rise to dissipation in the{ir} hydrodynamics.  Such dissipation terms appear as Navier-Stokes (NS) corrections to the GHD equation and was derived by Spohn \cite{spohn2012large}, Boldrighini and Suhov \cite{boldrighini1997one} and recently re-derived by Doyon and Spohn \cite{doyon2017dynamics}, and Ferrari and Olla \cite{ferrari2023macroscopic}. A simple scenario where the effect of this dissipation term gets  manifested is  the spreading of the position of a tagged (test) {quasi}particle (starting from a fixed position). Such a spreading was computed by  Lebowitz, Percus and Sykes (LPS) who demonstrated the effect of dissipation through microscopic calculations \cite{lebowitz1968time}. {It should be n}ote{d that} the NS term vanishes in the non-interacting limit of hard rods ($a \to 0$) {\it i.e.,} for hard point particles, because the collisions between two particles occur at a point and consequently, {the quasiparticles} cannot jump in space at collisions.  One generally expects that the presence of the dissipation terms makes the hard rod gas approach GGE.  The approach to GGE in hard rod system has recently been addressed by studying evolution of certain types of non-equilibrium initial states \cite{singh2024thermalization}. This study indicated that initial states with singular component in the velocity distribution does not seem to reach GGE and in such cases the solutions of  the Euler GHD equations displays most differences with molecular simulation at the locations of the shocks (in the Euler solutions). In \cite{singh2024thermalization},  such discrepancy was attributed to the presence of the Navier-Stokes corrections to the Euler GHD equation. 

In this paper, using a microscopic approach we study the evolution of hard rod system starting from generic non-equilibrium initial states.  
As mentioned earlier, a {hard rod} system has extensive number of {local} conserved quantities {denoted by $\{Q_\alpha\}$.} We study how the conserved densities evolve with time for a finite-sized  system  using a microscopic approach. While such expressions of the conserved densities are interesting, it is necessary to obtain the solution of the Euler  GHD equation from this microscopic solutions. We demonstrate how one can achieve this in the thermodynamic limit. 

The evolution of the conserved densities are often described well by the Euler solutions. However, there are cases (like the one mentioned above) in which Euler solution is not accurate enough. Secondly, the Boltzmann entropy \cite{chakraborti2022entropy, pandey2023boltzmann} of a system does not grow on the Euler solutions, although the system reaches a stationary state (often the GGE for integrable system) which has higher entropy than the initial state. Hence it is important to find solutions of the GHD equations beyond the Euler solutions. We aim to understand when and where the microscopic solutions provide improvements beyond Euler solutions. We apply our microscopic approach to study two particular  initial conditions studied previously: (a) Motion of a velocity tagged rod moving in the background of other rods. 
(b) Evolution of a domain wall initial condition in which rods with different velocity distributions are separated by a wall. 

Most of the studies in the literature concern annealed initial conditions, which are essentially chosen from an ensemble of initial conditions corresponding to some inhomogeneous macroscopic particle density profiles. In this study we also generalise some of our results to quenched initial conditions, in which the positions of the rods are fixed (quenched) initially.

The paper is organised as follows. We start with describing the conserved quantities and conserved densities of the hard rods in Sec.~\ref{sec:defn-ics-sum}. These are our quantities of interest to understand the macroscopic evolution of the system. This is followed by a discussion on the mapping to point particles and how one can choose initial conditions in the point particle picture and transform them to hard rod configurations. We end this section by providing a summary of our results. In Sec. \ref{sec:euler-GHD-and-NS}, we discuss the Euler GHD description of hard rod fluids and the possible Navier-Stokes corrections. We present the formalism and computation of the phase space density using a microscopic point of view in Sec.~\ref{sec:micro-calc-f(X,V,t)} for arbitrary annealed and quenched initial conditions. In this section we show how one can get the Euler solutions from the microscopic solutions in the thermodynamic limit. We also discuss about the possible correction to the Euler solution. The formulation presented in Sec.~\ref{sec:micro-calc-f(X,V,t)} is applied to two particular problems: (i) tracer diffusion in the background of hard rods in Sec.~\ref{sec:tracer-dyna} and (ii) the evolution of the interface separating two species of hard rod fluids having different velocity distributions in Sec.~\ref{sec:domain-wall}. {In Sec.~\ref{sec:NS-solution} we present numerical solution of the GHD equation with NS terms and compare with both the Euler solution and the microscopic solution.} Finally, in Sec.~\ref{sec:conclusion} we conclude our paper with possible future directions to explore. Some details of the calculations are provided in the appendix.

\section{Conserved densities, initial conditions and summary of results} 
\label{sec:defn-ics-sum}
In terms of the phase space density $f(X,V,t)$  defined in Eq.~\eqref{def:f(X,V,t)}, the conserved quantities $Q_\alpha=\sum_{i=1}^N V_i^\alpha$ for $\alpha=1,2,...,N$ can be re-written as  
\begin{align}
Q_\alpha(t) = \int dX \int dV V^\alpha f(X,V,t). \label{Q_a-rewrite}
\end{align}
From this equation, it is easy to identify the corresponding densities, $q_\alpha (X,t)$ of the conserved quantities
\begin{align}
Q_\alpha(t) = \int dX~q_\alpha(X,t),~~\text{where}~~q_\alpha(X,t) =  \int dV V^\alpha f(X,V,t). \label{def:q_alpha}
\end{align}
Note that $\alpha=0,~1$ and $2$ correspond to, respectively, the mass, the momentum and the energy density. The associated density profiles  $q_\alpha(X,t)$ can be  computed from the phase space density $f(X,V,t)$ for which, as we show in this paper, one can find an exact expression for certain types of initial conditions.

To do so, we first note that each configuration of the rods $\{X_i,V_i\}$ can be mapped to a configuration $\{x_i,v_i\}$ of hard point particles \cite{percus1969exact,bernstein1988expansion,lebowitz1968time}. The mapping is obtained by removing the non-accessible spaces between each pair of successive rods and is given by 
\begin{align}
x_i = X_i - \left(i - 1\right)a,~~\text{and}~~v_i=V_i~~\text{for}~i=1,2,...,N.~~~\text{(on infinite line)} \label{map-to-hpg-line}
\end{align}
If, instead of an infinite line, the rods are moving inside a box between two fixed hard walls at $X=0$ and $X=L$ then the above transformation changes to 
\begin{align}
x_i = X_i - \left(i - \frac{1}{2}\right)a,~~\text{and}~~v_i=V_i~~\text{for}~i=1,2,...,N.~~~\text{(in a box)}  \label{map-to-hpg-box}
\end{align}
The positions of the walls at $X=0$ and $X=L$ in the hard rod (HR) picture get transformed to, respectively, the wall at $x=0$ and $x=L-Na$ in the hard point particle (PP) picture. This means the hard point particles move inside a smaller box of size $L'=L-Na$. The hard rods collide with the walls of the box elastically, such that at each collision, their velocities get reversed. Same elastic collisions with the walls happen in the point particle picture as well. Hence, corresponding to a collection of hard rods, there is an equivalent hard point gas (HPG).

We are interested to find the single particle phase space distribution defined in Eq.~\eqref{def:f(X,V,t)} where the average $\langle \cdots \rangle$ is performed over initial configurations $\{X_i,V_i\}$ chosen from an ensemble. Since for each configuration $\{X_i,V_i\}$ of hard rods, there is a unique configuration $\{x_i,v_i\}$ of hard point particles, one can specify the ensemble of initial configurations in terms of the HPG configurations and define the following single particle phase space distribution for hard point particles, {or, equivalently the quasiparticles in the PP picture}  (denoted by superscript `0' )
 \begin{align}
 f^0(x,v,t)=\sum_{i=1}^N\langle\delta(x-x_i)\delta(v-v_i)\rangle. \label{def:f^0(x,v,t)}   
\end{align} 
It is easy to realise that the single particle phase space density $f^0(x,v,t)$ satisfies the following hydrodynamic (HD) equation
\begin{align}
\partial_t f^0(x,v,t) +v\partial_xf^0(x,v,t) = 0, \label{hd-eq-f^0}
\end{align}
which is essentially the Liouville equation in the phase space.
\noindent
In this paper we consider two types of initial ensembles. In each of them, the initial configuration is chosen in PP picture and then transformed to the HR picture. The ensembles are specified as follows:
 \begin{itemize}
 \item[1.] Annealed initial conditions (AIC): In the {PP} picture, position{s} and velocit{ies} of individual particles are chosen independently and identically from the distribution $\mu_a(x,u)$. The joint distribution of the positions and momenta of the particles, in this case, is  
 \begin{align}
 \mathbb{P}_{a}(\{x_{i},u_{i}\},0)=\prod_{i=1}^N \mu_{a}(x_{i},u_{i})~\prod_{i=1}^{N-1}\Theta(x_{i+1}-x_i), \label{en-aa}
 \end{align}
where $\Theta(x)$ is the Heaviside theta function. Note that the product over $\Theta$ functions ensures the ordering $\{x_{i}\le x_{i+1}~;~i=1,2,...,N\}$.
The marginal probability densities and the conditional velocity distribution are given by
\begin{align}
p_{a}(x) =\int du ~\mu_{a}(x,u),~~h_{a}(u)=\int dx~ \mu_{a}(x,u),~~h_{a}(u|x) = \frac{\mu_{a}(x,u)}{p_a(x)}. \label{def:p_a-h_a}
\end{align}
Hence the initial phase space density of the point particles is  $f^0(x,u,0)= N \mu_{a}(x,u)$ such that $\int dx \int du f^0(x,u,0)=N$. For specified macroscopic density profiles of the conserved quantities such as mass, momenta, energy, etc., one can choose the distribution $\mu_a(x,u)$  appropriately. {For simplicity, in this paper we have mostly considered product distribution $\mu_a(x,u) = p_a(x)h(u)$ except in Sec.~\ref{sec:domain-wall}.}

 \item[2.] Quenched initial conditions (QIC):  In this case the positions of the particles are fixed $\{\bar{x}_{i}~|~\bar{x}_i \le \bar{x}_{i+1}~;~i=1,2,...,N\}$ at $t=0$ for each realisation, however, the velocities are chosen independently from some distribution $h_{q}(v|\bar{x})$. The joint distribution of the positions and momenta of the particles can be written as 
 \begin{align}
  \mathbb{P}_{q}(\{x_{i},u_{i}\},0)=\prod_{i=1}^N \delta(x_{i}-\bar{x}_{i})~h_{q}(u_{i}|\bar{x}_{i}). \label{en-qa}
 \end{align}
 The corresponding phase space density at $t=0$ is then given by $ f^0(x,u,0)=\sum_i \delta(x-\bar{x}_{i})h_{q}(u|\bar{x}_{i})$ which also normalises to $N$. In this case, one can also arrange the initial positions $\bar{x}_i$ according to a given mass density profile {$\varphi_q(\bar{x}) = \int du f^0(x,u,0)$} and choose the distribution $h_q(u|x)$ according to given momentum and energy profiles. {Once again for simplicity, we have chosen $h_q(u|x) =h(u)$ to be independent of $x$. Our results for quenched initial configurations can be straightforwardly extended to the case of space dependent velocity distribution. }
 \end{itemize}
 Once a position and velocity configuration $\{x_i,v_i\}$ is chosen in PP picture, we convert this to hard rod configurations using the mapping
 \begin{align}
 X_i = x_i + \left(i - 1\right)a,~~\text{and}~~V_i=v_i~~\text{for}~i=1,2,...,N.~~~\text{(on infinite line)}. 
  \label{inv-map-to-hpg-line}   
 \end{align}
For configurations inside a box, we use the mapping in Eq.~\eqref{map-to-hpg-box}. These initial conditions are then evolved to get the final configuration at time $t$, and the quantities of interest such as the conserved density profiles are computed from the final configurations. 

{The goal of this paper is to study the evolution of a collection of hard rods on a line microscopically to find the time evolution of conserved densities on macroscopic space-time scale. Such macro-evolution are usually governed by the solutions of generalised hydrodynamic equations and it is interesting to see how such HD solutions appear from microscopic computations. Derivation of Euler solutions from microscopic dynamics has been carried out previously in\cite{percus1969exact,lebowitz1968time,bernstein1988expansion,jepsen1965dynamics,valleau1970time}, however mostly for factorised and annealed initial conditions. We present a rigorous and clear microscopic approach which, in addition to reproducing the previous results, enables us to extend the study to more general initial conditions such as quenched initial conditions and non-factorizable (position dependent velocity distribution) annealed initial conditions. We also demonstrate the deviation from the Euler solutions both using  microscopic analysis and (numerical) solution of the GHD equation with Navier-Stokes correction. Below we provide a point-wise summary of our results.}
\begin{itemize}
\item[(i)] Through microscopic computations, we find exact expressions for the density profiles of the conserved quantities at any time $t$ for arbitrary number, $N$ of hard rods. We study both annealed and quenched initial conditions.

\item[(ii)] We demonstrate how taking the thermodynamic limit of the microscopic solutions appropriately, one finds the solutions of the Euler GHD. We find that the Euler solutions, corresponding to same initial conserved density profiles are same for both quenched and annealed initial conditions. However, the solutions beyond the Euler ones are different in the two cases. {In the problem of tracer rod motion, we explicitly show that the variance of the tracer position is smaller in the quenched case  compared to the annealed case.}

\item[(iii)] To demonstrate this more quantitatively we revisit the problem of tracer rod evolution in the background of other rods. This problem was studied previously by LPS in ~\cite{lebowitz1968time} where they had shown that the tracer rod performs a drift-diffusion. 
In this paper, we provide an alternative microscopic derivation of the solution of the tracer rod diffusion for both quenched and annealed initial conditions. 
In addition to reproducing the LPS results for homogeneous background, we generalise their results for in-homogeneous backgrounds as well.

\item[(iv)] {We find the microscopic solution for a particular domain wall initial condition in which an interface (wall) in the middle separates two 
different species of {hard rod} gas with different velocity distributions. This initial condition was studied recently in ~\cite{singh2024thermalization} where 
it was shown that the Euler solution of the density profiles of each species has a jump at the interface (wall) which moves ballistically with some effective speed. However, the Euler solution does not agree with molecular dynamics  at the location of the jump (or shock). In fact the numerical profiles displayed broadening of the interface at the shock location. We study this broadening in Sec.~\ref{sec:domain-wall} and Sec.~\ref{sec:NS-solution}, respectively,  using microscopic approach and solving the GHD equation with NS correction numerically.}
 
\end{itemize}

\section{The Euler GHD equation and the NS correction}
\label{sec:euler-GHD-and-NS}
The GHD description of hard rods given in Eq.~\eqref{eq:GHD} is a hydrodynamic equation at Euler scale (ballistic scale). Its solution has been discussed in different previous works \cite{doyonlecturenotes, doyon2017dynamics, singh2024thermalization}. For completeness, we here briefly discuss the solution as presented in \cite{singh2024thermalization}. The idea for solving the Euler GHD equation is to transform it into the GHD equation for hard point particle as in Eq.~\eqref{hd-eq-f^0}. Since each configuration of hard rods can be mapped to a unique configuration of hard point particles, there should be a relation between the single particle phase space density of hard rods and that of point particles. If there are $n$ hard rods on a phase space region $\Delta X \Delta V$ around a phase space point $(X,V)$, then under the transformation in Eq.~\eqref{map-to-hpg-line} these rods would map to $n$ point particles on a smaller phase space region $\delta x \delta v$ around a phase space point $(x,v)$ such that $x=X-na$ and $v=V$. Similarly, $\delta x=\Delta X - na$ and $\Delta V=\delta v$. Hence the {phase space distribution of point particles 
\begin{align}
f^0(x(X),v,t) &= \frac{n}{\delta x \delta v} = \frac{n}{(\Delta X - na) \Delta V} = \frac{f(X,V,t)}{1-a \rho(X,t)}, \label{f-f^0-rela}
\end{align}
{where}
\begin{align}
x(X)=X-a F(X,t),~~v=V, \label{hd-mapping} 
\end{align}
{and} the mass density, {$\rho (X,t)$} and the cumulative density, {$F (X,t)$} are given by 
\begin{align}
\rho(X,t)=\int dV~f(X,V,t),~\text{and}~F(X,t) = \int^XdY \rho(Y,t). \label{def:rho-F}
\end{align}
}
Note that $\rho(X,t)$ represents the mass density of hard rods defined in Eq.~\eqref{def:rho-U} and $F(X,t)$ is the cumulative mass till point $X$ at time $t$. Also note that the transformation $x=X-a F(X,t)$ is in fact  same as the transformation in Eq.~\eqref{map-to-hpg-line} except now it is written in terms of hydrodynamic density field. The term $-aF(X,t)$ essentially removes the unaccessible space between rods below position $X$ which is represented by $a$ times the number of rods $F(X,t)$ below $X$. 

Next task is to convert the GHD equation \eqref{eq:GHD} into the GHD equation of point particles in Eq.~\eqref{hd-eq-f^0} using  the relations in Eqs.~\eqref{f-f^0-rela} and \eqref{hd-mapping}. For this we define the normal density {(denoted by the subscript `$n$')}
\begin{align}
f_n(X,V,t) = \frac{f(X,V,t)}{1-a \rho(X,t)}, \label{def:f_n}
\end{align}
where $\rho(X,t)$ is defined in Eq.~\eqref{def:rho-U}.
Using Eq.~\eqref{f-f^0-rela} in Eq.~\eqref{def:f_n} we observe
\begin{align}
f_n(X,V,t)=f^0\left(X-a F(X,t),V,t\right). \label{f_n-f^0-rela}
\end{align}
Furthermore, using $\partial_t \rho(X,t) = -\partial_X(\rho U)$, we also find that the normal distribution function satisfies 
\begin{align}
\partial_t f_n(X,V,t) +V_{\rm eff}(X,t) \partial_X f_n(X,V,t)=0, \label{f_n-eq}
\end{align}
where the expression of $V_{\rm eff}(X,t)$ is given in Eq.~\eqref{def:V_eff}.  Now inserting the relation between distributions $f_n$ and $f^0$  in Eq.~\eqref{f_n-f^0-rela} and simplifying further using the relation $\partial_tF(X,t)+\rho U=0$, one finds Eq.~\eqref{hd-eq-f^0} which is very easy to solve. Given initial condition $f^0(x,v,0)$, the solution at time $t$ is obtained just by translating the function by amount $-vt$ {\it i.e.,} $f^0(x,v,t) =f^0(x-vt,v,0)$. Once the solution at time $t$ is known in point particle (PP) picture, one transforms back to the hard rod (HR) picture by the inverse 
transformation \cite{singh2024thermalization}
{
\begin{align}
f(X(x),V,t) = \frac{f^0(x,V,t)}{1+a\rho^0(x,t)}, \label{sol:euler-GHD-hr}
\end{align}
with $X(x)=x+a F^0(x,t)$ where $ \rho^0(x,t)=\int dv f^0(x,v,t)$ is the mass density of point particles and $F^0(x,t)=\int^xdy~\rho^0(y,t)$ is the corresponding cumulative density.}
It has been shown in \cite{doyon2017dynamics, singh2024thermalization} that, in the hydrodynamic limit, the solution in Eq.~\eqref{sol:euler-GHD-hr} describes the molecular dynamics simulation quite well for generic initial conditions. However, as demonstrated in \cite{singh2024thermalization}, for certain initial conditions the solution of Euler GHD [in Eq.~\eqref{sol:euler-GHD-hr}] does not match with molecular dynamics simulation. It was pointed out and argued that these discrepancies arise from the dissipation terms in the GHD \cite{singh2024thermalization} which appear as Navier-Stokes (NS) corrections to Euler GHD. Such a correction term for {hard rod} systems was first computed by Spohn \cite{spohn2012large}. Later, Boldrighini and Suhov  established the NS term through rigorous calculation \cite{boldrighini1997one}. Recently, Doyon and Spohn re-derived the NS term using a Green-Kubo calculation \cite{doyon2017dynamics}, and Ferrari and Olla  \cite{ferrari2023macroscopic} re-established the form of NS in a more general setting through rigorous calculations. A simpler way of getting the NS term and possible modifications are discussed recently, in the context of revising Enskog equation for hard rod gas \cite{bulchandani2024revised}. Very recently, a possible correction and new form of the Navier-Stokes term has been proposed \cite{hubner2024diffusive}.
With the NS correction, the GHD equation modifies to \cite{doyon2017dynamics, singh2024thermalization}
\begin{align}
&\partial_t f(X,V,t) + \partial_X V_{\rm eff}(X,V,t) f(X,V,t) = \partial_X \mathcal{N}(X,V,t),~~\text{where,} \label{eq:GHD-NS} \\
& \mathcal{N}(X,V,t) = a^2\int dW|V-W|~\frac{
\left( f(X,W,t)\partial_Xf(X,V,t) - f(X,V,t)\partial_Xf(X,W,t) \right)}{2(1-a\rho(X,t))}. 
\nonumber
\end{align}
In general, it seems hard to solve Eq.~\eqref{eq:GHD-NS}. In the next section we investigate if one can find an expression for $f(X,V,t)$ from microscopic calculation which goes beyond the Euler solution in Eq.~\eqref{sol:euler-GHD-hr}.

\section{Phase space density $f(X,V,t)$ from microscopic calculation}
\label{sec:micro-calc-f(X,V,t)}
In this section we provide exact microscopic computation of the single particle phase space {distribution function $f(X,V,t)$} for finite number, $N$ of rods moving on a line.

We start by noting that the distribution $f(X.V,t)$ in Eq.~\eqref{def:f(X,V,t)} can be written as 
{
\begin{subequations}
\begin{equation}
f(X,V,t) = \sum_{i=1}^N {\bf P}_i(X,V,t), \label{f-P_i-rel}
\end{equation}
\text{where}
\begin{equation}
{\bf P}_i(X,V,t) = \left \langle \delta\left (X-X_i(t)\right ) \delta(V-V_i) \right \rangle, \label{f-P_i-rel-1}
\end{equation}
\end{subequations}
}
represents the probability (density) of finding the $i^{\rm th}$ (counted from the left) hard rod at phase space point $(X,V)$ at time $t$. 
Performing the mapping to the HPG coordinates from  Eq.~\eqref{map-to-hpg-line} in the above equation, we can write $f(X,V,t)$ as 
{
\begin{subequations}
 \label{def:x*_i} 
\begin{equation}
f(X,V,t) = \sum_{i=1}^N  \mathtt{P}_i(x_i^*,V,t), 
\end{equation}
\text{with}
\begin{equation}
x_i^*=X-(i-1)a,
\end{equation}
\text{where,}
\begin{equation}
\mathtt{P}_i(x,v,t)= \left \langle \delta \left (x_i(t) -x\right ) \delta(v-v_i) \right \rangle, 
\end{equation}
\end{subequations} }
represents the probability (density) of finding the $i^{\rm th}$ point particle at position $x$ with velocity $v$ at time $t$. 
To compute the probability distribution $\mathtt{P}_i(x,v,t)$, we start with the $N$ particle propagator of $N$ hard point particles 
\begin{align}
\mathbb{G}_N\left({\bf y},{\bf v},t|{\bf x},{\bf u},0\right)=\sum_{\tau \in \sigma_N} \prod_{k=1}^N g(y_k,v_k,t|x_{\tau(k)},u_{\tau(k)},0),
\end{align}
where ${\bf y}=(y_1,y_2,...,y_N)$,  ${\bf v}=(v_1,v_2,...,v_N)$, ${\bf x}=(x_1,x_2,...,x_N)$ and ${\bf u}=(u_1,u_2,...,u_N)$. {Here, $y_{i} \leq y_{i+1}$ and $x_{i} \leq x_{i+1}$ for $i = 1, \cdots, N-1$. } The set $\sigma_N$ in the above equation represents the set of all permutations of $N$ particle labels, $\tau$ is an element of this set  and $\{x_i,u_i\}$ are initial set of positions and velocities. The function $g(y,v,t|x,u,0)$ represents the single particle propagator for reaching the phase space point $(y,v)$ at time $t$ starting from phase space point $(x,u)$. Although for ballistic hard point particles, this propagator can be formally written as  
\begin{align}
g(y,v,t|x,u,0) = \delta(y-x-ut)\delta(v-u), \label{prop-sp-ballistic}
\end{align}
we, however, present the calculation for general single particle propagator.

Let $\mathscr{P}_i(z,v,t|{\bf x},{\bf u})  $ represent the probability of finding the $i^{\rm th}$ point particle at position $z$ with velocity $v$ at time $t$ given that initial positions and the velocities of the particles are ${\bf x}$ and ${\bf u}$. This probability can be obtained by integrating the $N$ particle propagator $\mathbb{G}_N({\bf y},{\bf v},t|{\bf x},{\bf u},0)$ over final positions and velocities of  all particles except those for the $i^{\rm th}$ particle which are fixed at $z$ and $v$, respectively. As shown in \ref{derivation of mttP(z,v,t)-line}, we find 
{
\begin{subequations}
\label{def:q_p-q-m} 
\begin{align}
\mathscr{P}_i(z,v,t|{\bf x},{\bf u})  
&=\sum_{m=1}^N \sum_{S_\ell, S_r}  g(z,v,t|x_m,u_m,0)~\prod_{k=1}^{i-1} g_{<}(z,t;x_{S_\ell[k]},u_{S_\ell[k]},0),  \\
&~~~~~~~~~~~~~~~~\times~~\prod_{k=i+1}^{N} g_{>}(z,t;x_{S_r[k]},u_{S_r[k]},0),  \nonumber
\end{align}
with
\begin{equation}
g_{<}(z,t|x,u,0)  = \int_{-\infty}^z dy\int_{-\infty}^\infty dv~g(y,v,t|x,u,0),
\end{equation}
and
\begin{equation}
g_{>}(z,t|x,u,0) =1-g_{<}(z,t|x,u,0), 
\end{equation} 
\end{subequations} }
where summations are done over all possible partitions of $N$ labels of the particles into three groups containing  $(i-1)$, one and $(N-i)$ particles, represented by set $S_\ell$, index $m$ and set $S_r$ respectively. The notation ${S_\ell[k]}$ represents the $k^{\rm th}$   element of the set of $(i-1)$ labels in the partition $S_\ell$. Similar meaning holds for  ${S_{r}[k]}$. 
In the next step one needs to average over the initial configurations $\{x_i,u_i\}$. At this point we present the discussions for the two types of initial conditions annealed [see Eq.~\eqref{en-aa}] and quenched [see Eq.~\eqref{en-qa}] separately.

\subsection{Annealed initial conditions:}
In this case, both the position and the velocity of the particles are random and chosen independently for each particle from a distribution $\mu_a(x,v)$. Averaging over this initial configuration, one finds [see \ref{ap:annealed-case} for details]
{
 \begin{align}
\mathtt{P}_i(z,v,t) &= \prod_{k=1}^N  \int dx_k \int du_k~\mathscr{P}_i(z,v,t|{\bf x},{\bf u})  ~ \prod_{j=1}^N \mu_{a}(x_j,u_j) \notag \\ 
&= N\binom{N-1}{i-1} q_a(z,t)^{i-1} ~g^0(z,v,t)~(1-q_a(z,t))^{N-i}, \label{mttP(z,v,t)} 
\end{align}
where,
\begin{align}
g^0(z,v,t) &= \int dx \int du~g(z,v,t|x,u,0)~\mu_{a}(x,u), \label{def:g^0-g} \\ 
p_a(z,t) &= \int dv \  g^0(z,v,t),~\text{and}~~q_a(z,t) = \int^z dy \ p_a(y,t). \label{def:q-gm}
\end{align} }
\begin{figure}[]
    \centering
    \includegraphics[scale=0.42]{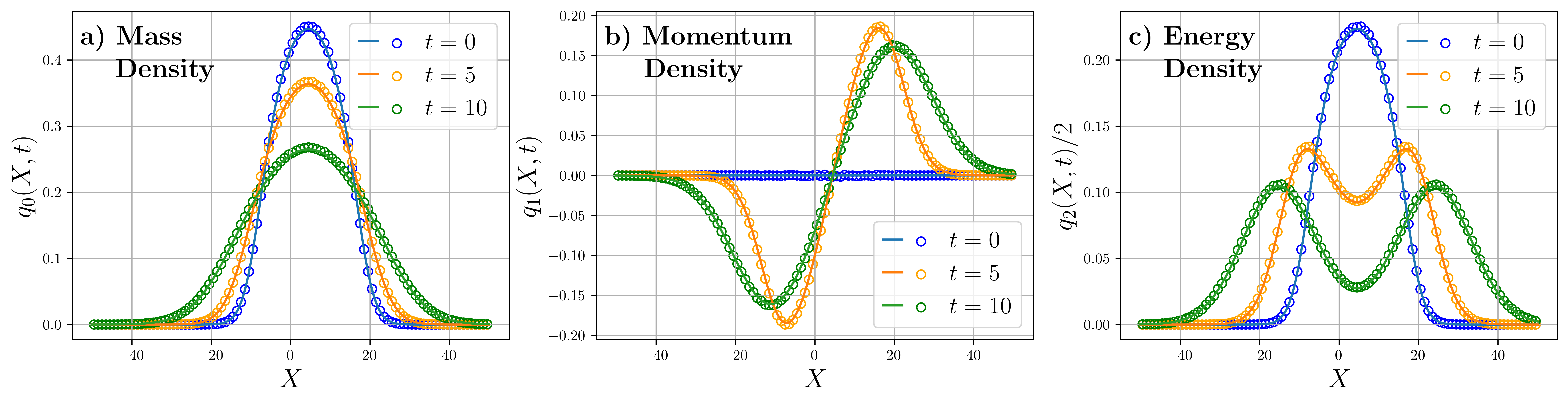}
    \caption{Plot showing comparison of the mass $q_0(X,t)$, momentum $q_1(X,t)$ and energy $\frac{1}{2}q_2(X,t)$ density profiles of $N=10$ hard rods at different times. The initial positions and velocities of the rods are chosen independently first in PP picture from Gaussian distribution as in Eq.~\eqref{eq:ic-gaussian} and Eq.~\eqref{eq:max-dist},  and then transformed to rod configurations using the mapping of Eq.~\eqref{inv-map-to-hpg-line}. The parameters of the plots are $a=1.0,~\sigma=5a$ and $T=1.0$. The circles represent simulation data obtained after averaging over $10^6$ realizations and solid lines represent theoretical result for $q_\alpha(X,t)$ obtained from Eq.~\eqref{def:q_alpha} using the phase space distribution in Eq.~\eqref{f(X,V,t)-on-line}.}
    \label{fig:profiles-gaussian}
\end{figure}
\noindent 
{To obtain the final expression in Eq.~\eqref{mttP(z,v,t)}, we have used the fact that the point particles are identical and their positions $\{x_i\}$ and velocities $\{u_i\}$ are chosen independently, making them iid variables. A more intuitive way to understand Eq~\eqref{mttP(z,v,t)} is as follows: We note that} the factor $g^{0}(z,v,t)$ represents the probability of placing a point particle at phase space  point $(z,v)$ at time $t$. This can be done in $N$ ways. The factors $q_a(z,t)^{i-1}$ and $(1-q_a(z,t))^{N-i}$ represent, the probability to put $(i-1)$ and $(N-i)$ particles, respectively, below and above the position $z$ and thus making the particle at $z$ to be the  $i^{\rm th}$ particle at time~$t$. The binomial factor represents, the number of ways to chose $(i-1)$ particles from the remaining $(N-1)$ particles.\\[3pt]
Inserting the expression of  
$\mathtt{P}_i(z,v,t)$ from Eq.~\eqref{mttP(z,v,t)} in Eq.~
\eqref{def:x*_i}, 
we get 
\begin{align}
\begin{split}
f(X,V,t) = \sum_{i=1}^N N \binom{N-1}{i-1} &\left[q_a\left(X-(i-1)a,t\right)\right]^{i-1} g^0\left(X-(i-1)a,V,t\right) \\ 
&~~~~~~~~~~\times~~
\left[1-q_a(X-(i-1)a,t)\right]^{N-i}.
\end{split}
\end{align}
The phase space density of point particles is given by  
\begin{align}
f^0(z,v,t)=Ng^0(z,v,t), \label{rel:f^0-p_a}
\end{align} 
such that the mass density and the cumulative density of the point particles are given by 
\begin{align}
\rho^0(z,t)&=Np_a(z,t), \label{den-pp}\\
F^0(z,t)&=Nq_a(z,t). \label{cu-den-pp}
\end{align}
In terms of $f^0(z,v,t)$ and $F^0(z,t)$ defined above, one can re-write $f(X,V,t)$ as  
\begin{align}
\begin{split}
f(X,V,t)
= \sum_{i=0}^{N-1} \binom{N-1}{i} &\left(\frac{F^0(X-ia,t)}{N} \right)^{i} ~f^0(X-ia,V,t) \\ 
&~~~~~~\times~~\left(1-\frac{F^0(X-ia,t)}{N}\right)^{N-1-i}, 
\end{split}
\label{f(X,V,t)-on-line}
\end{align}
using which in Eq.~\eqref{def:q_alpha}, one can easily compute the density profiles of other conserved quantities, $q_\alpha(X,t)$ for $\alpha=1,2,...,N$. Inserting this expression of $f(X,V,t)$ in Eq.~\eqref{def:q_alpha} and performing the integral over $V$, one can now find profiles of the conserved densities, such as mass, momenta and energy. In Fig.~\ref{fig:profiles-gaussian} we compare the mass, momenta and energy profiles for $N=10$ hard rods obtained theoretically using Eq.~\eqref{f(X,V,t)-on-line} with the same obtained from numerical simulation. Initial conditions for the rods are chosen in the PP picture [see eq.~\eqref{en-aa}] with 
\begin{align}
\mu_a(x,v) &= \frac{\exp\left(-\frac{x^2}{2 \sigma^2}\right)}{\sqrt{2 \pi \sigma^2}}~\times~h(v), \label{eq:ic-gaussian} \\ 
\text{where}~&~h(v)=\frac{\exp\left(-\frac{v^2}{2 T}\right)}{\sqrt{2 \pi T}}. 
\label{eq:max-dist}
\end{align}
and then transformed to rod configurations using the mapping in Eq.~\eqref{inv-map-to-hpg-line}. For all three profiles we observe excellent agreement between theory and simulation.

{\color{black}
In order to take the large $N$ limit, often it seems convenient to go to the Fourier space. We define 
\begin{align}
\hat{f}(k,V,t) = \int_{-\infty}^\infty dX~e^{\iota k X} f(X,V,t),\label{def:fhat}
\end{align}
such that 
\begin{align}
f(X,V,t) = \int_{-\infty}^\infty \frac{dk}{2\pi}~e^{-\iota k X} \hat{f}(k,V,t),\label{def:fhat-inv}
\end{align}
Inserting the expression of $f(X,V,t)$ from Eq.~\eqref{f(X,V,t)-on-line} in Eq.~\eqref{def:fhat} and simplifying, one gets 
\begin{align}
\hat{f}(k,V,t) =  \int_{-\infty}^\infty dz ~e^{\iota k z}~ \left [ 1 + \left(e^{\iota k a}-1\right) \frac{F^0(z,t)}{N} \right ]^{N-1} f^0(z,V,t), 
\label{fhat}
\end{align}
where recall $F^0(z,t)=Nq_a(z,t)$ from Eq.~\eqref{cu-den-pp}. The Eq.~\eqref{fhat} is still exact for any finite $N$ and $a$. To get the distribution function in the hydrodynamic limit, we now take the $N \to \infty$ and $a \to 0$ limit, keeping $Na$ fixed.
 For that, we approximate  $e^{\iota k a}= 1+\iota k a + O(k^2a^2)$, and then taking the above limits, keeping $Na$ fixed we get, 
\begin{align}
\hat{f}(k,V,t) \approx \int_{-\infty}^\infty dz ~e^{\iota k (z+ a F^0(z,t))}f^0(z,V,t). \label{fhat-eu}
\end{align}
which upon performing the inverse Fourier transform in Eq.~\eqref{def:fhat-inv} immediately yields, 
\begin{align}
f(X,V,t) &= \int_{-\infty}^\infty dz~\delta\left(X-z- aF^0(z,t)\right)~f^0(z,V,t) = \frac{f^0(z^*(X),V,t)}{1+a \rho^0(z^*(X),t)}, 
\label{sol:euler-micro}
\end{align}
where $z^*(X)$ is the solution of the equation
\begin{align}
X=z^* + a F^0(z^*,t). \label{hd-transfrmtn}
\end{align}
The Eq.~\eqref{sol:euler-micro} along with Eq.~\eqref{hd-transfrmtn} provides the solution of the Euler GHD equation obtained in Eq.~\eqref{sol:euler-GHD-hr} and was also obtained previously in \cite{percus1969exact}. Note that,  This solution approximates the microscopic density profile in Eq.~\eqref{f(X,V,t)-on-line} on the largest space-time scales in the hydrodynamic limit. Keeping higher order terms in the expansion $e^{\iota k a} = 1+\iota k a -k^2a^2/2 + ...$, one can introduce finer structures in the distribution function at smaller and smaller length scales.  Keeping terms of all powers in $a$, one can simplify the Eq.~\eqref{fhat} in the $N \to \infty$ limit as 
\begin{align}
\hat{f}(k,V,t) =  \int_{-\infty}^\infty dz ~e^{\iota k z}~ \exp \left [ \left(e^{\iota k a}-1\right)F^0(z,t) \right ] f^0(z,V,t).
\label{fhat-Ninf}
\end{align}
which on performing inverse Fourier transformation provides the following phase-space density of the hard rods 
\begin{align}
f(X,V,t) &=\sum_{i=0}^{\infty} \exp\left[-F^0(X-ia,t)\right] ~\frac{\left[F^0(X-ia,t) \right]^{i}}{i!} ~f^0(X-ia,V,t),
\label{f(X,V,t)-N-to-inf}
\end{align}
as was obtained previously by Percus \cite{percus1969exact}. Note, the expression in Eq.~\eqref{f(X,V,t)-N-to-inf} can also be obtained directly from Eq.~\eqref{f(X,V,t)-on-line} by first, using Stirling approximation for the factorials involving $N$ and then taking the $N \to \infty$ limit at fixed $X$ and $t$. It is easy to understand the different terms in the expression of $f(X,V,t)$ in Eq.~\eqref{f(X,V,t)-N-to-inf}. There are different ways (events) to find a rod  at location $X$ at time $t$ with velocity $V$. The factor $f^0(X-ia,V,t)$ represents the probability density of finding a point particle at position $x=X-ia$ at time $t$ with velocity $V$. The other two factors together represent the Poisson distribution of finding $i$ number of point particles below the position $x=X-ia$. Clearly such a configuration of point particles would contribute to the probability density of finding a hard rod at location $X$ at time $t$ with velocity $V$. Adding contributions from all such events gives rise to the distribution $f(X,V,t)$ as present in Eq.~\eqref{f(X,V,t)-N-to-inf}. 

In order to get an approximate expression of the distribution function for large but finite $N$, we first rewrite the $[...]^{(N-1)}$ term in Eq.~\eqref{fhat} as 
$\exp \left((N-1)\log \left[ 1+ \left(e^{\iota k a}-1\right) \frac{F^0(z,t)}{N}\right] \right), $
and expand in powers of $1/N$ and $a$ for large $N$ and small $a$ keeping $Na$ finite. We get 
\begin{align}
\hat{f}(k,V,t) \approx  \int_{-\infty}^\infty dz ~e^{\iota k z}~ \exp \left [ \iota k a F^0(z,t) - \frac{k^2a^2}{2N}F^0(z,t)(N-F^0(z,t))\right ] f^0(z,V,t).
\label{fhat-Ninf-2}
\end{align}
Performing inverse Fourier transformation one finds,  
\begin{align}
f(X,V,t) 
&\approx \int_{-\infty}^\infty dz \frac{\sqrt{N}\exp \left( - \frac{N(X-z-aF^0(z,t))^2}{2 a^2\Sigma_a^2(z,t)}~ \right)}{\sqrt{2 \pi a^2 \Sigma_a^2(z,t)}}~f^0(z,V,t), 
\label{sol:beyond-Euler-2} 
\end{align}
where 
\begin{align}
\Sigma_a^2(z,t) = F^0(z,t) (N - F^0(z,t))=NF^0(z,t) -\left( \int dx~ \varphi_a(x) g_<(z,t|x,0)\right)^2, \label{Sigma_a^2}
\end{align}
with $\varphi_a(x) = \rho^0(x,0)=N p_a(x)$ [see Eq.~\eqref{def:p_a-h_a} and also Eq.~\eqref{den-pp}]. 
The  Eq.~\eqref{sol:beyond-Euler-2} essentially provides a expression for the macroscopic distribution function $f(X,V,t)$ that goes beyond the Euler solution in Eq.~\eqref{sol:euler-micro} where the delta function inside the integral got replaced by a Gaussian substructure. Similar correction with Gaussian substructure was presented in \cite{percus1969exact}, but in the $N \to \infty$ limit. The Eq.~\eqref{sol:beyond-Euler-2} provides a large (but finite) $N$ version of similar correction. 
It is easy to see that in the $N\to \infty$ and $a \to 0$ limit, keeping $Na$ fixed, one can perform the integral in Eq.~\eqref{sol:beyond-Euler-2} by saddle point method and once again one recovers the Euler solution in Eq.~\eqref{sol:euler-micro} as one should.
}

\begin{figure}[]
    \centering
    \includegraphics[scale=0.42]{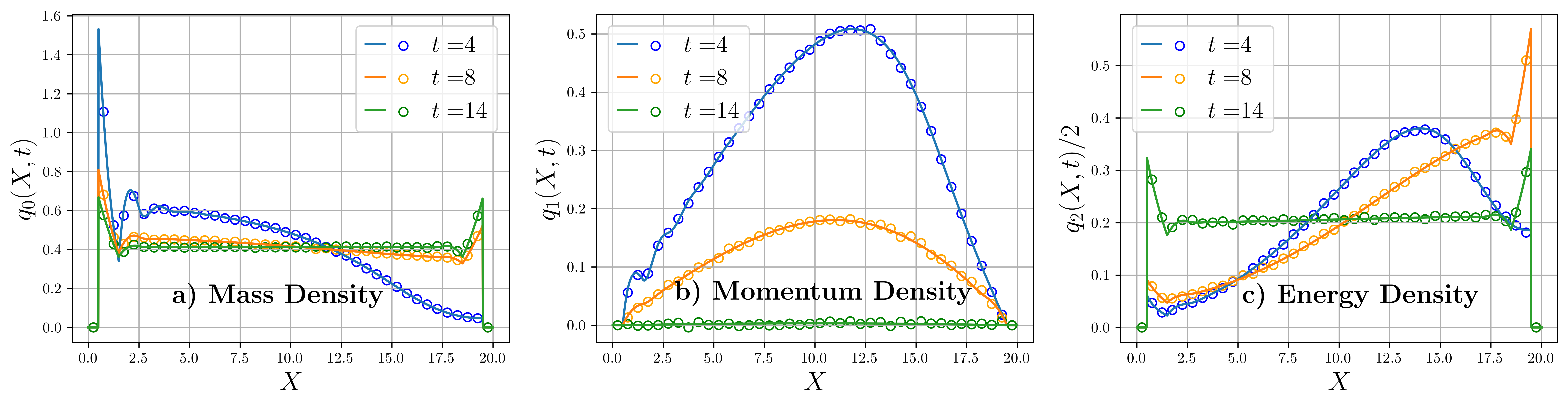}
    \caption{Plot showing comparison of the mass $q_0(X,t)$, momentum $q_1(X,t)$ and energy $\frac{q_2(X,t)}{2}$ density profiles of $N=8$ hard rods inside a box of length $L=20$ at different times. The initial configuration is chosen in the PP picture by choosing their individual positions in the left half $[0,(L-Na)/2]$ of the box with uniform probability, and choosing their individual velocities from Maxwell distribution $h(v)$ in Eq.\eqref{eq:max-dist}. The parameters of the plots are $a=1.0$ and $T=1.0$. The circles represent simulation data obtained after averaging over $10^5$ realizations and solid lines represent theory.}
    \label{fig:profiles-box-fe}
\end{figure}

\paragraph{Rods moving inside a box of length $L$:}

The above discussion for the single particle phase space density of rods corresponds to the case in which they are moving on an infinite straight line. One can generalise the steps of the derivation to the case when the rods are restricted to move inside a box $[0,L]$ of length $L$.  In this case one finds that the phase space density $f(X,V,t)$ is given by 
{
\begin{align}
&f(X,V,t) =  \sum_{i=1}^N \mathtt{P}_i(x_i^*,V,t)~\Omega_i[X,L,N],  \label{def:f(X,V,t)-tran-2} 
\end{align}
where $x^*_i=[X-(i-1/2)a] $, and 
\begin{align}
\Omega_i[X,L,N]=\Theta \left[X-a \left(i-\frac{1}{2}\right)\right] \Theta 
\left[L-a\left(N-i+\frac{1}{2}\right)-X\right], \label{Theta} 
\end{align} }
and $\mathtt{P}_i(z,v,t)$ is given in Eq.~\eqref{mttP(z,v,t)}. The function $\Omega_i[X,L,N]$ involving Heaviside theta function $\Theta(z)$ indicates that the maximum available region of space for the $i^{\rm th}$ hard rod inside the box is restricted and it is given by $(i-1/2)a < X <L-a\left(N-i+1/2\right)$. This corresponds to the configuration in which $(i-1)$ rods are tightly stuck to the left wall  while the remaining $(N-i)$ rods are stuck to the right wall of the box. In Fig.~\ref{fig:profiles-box-fe}, we plot the profiles of the mass, momentum and energy density profiles of $N=8$ hard rods at three different times. Initially, the positions of the $N=8$ point particles are chosen in the left half of the box with uniform probability  and each of them given a random velocity chosen independently from Maxwell velocity distribution $h(v)$ in Eq.~\eqref{eq:max-dist} at temperature $T=1.0$. 
To compute these profiles, we perform event-driven simulation in which we first evolve each point particle configuration to the final time and then convert the resultant configuration to hard rod configurations using the mapping in Eq.~\eqref{map-to-hpg-box}. Now in this hard rod configuration we compute the conserved densities and then average over many initial conditions. We observe excellent agreement of the numerical data with the analytical expressions. We also observe that the momentum of the system finally goes to zero as momentum conservation gets lost at the collisions of the rods with the boundaries of the box. We find  that the other two density profiles approach to the stationary values (which is homogeneous in the bulk) at large time limit. 

 \subsection{Position quenched  initial conditions:}
 In this case, the initial positions $\{\bar{x}_k\}$ of the particles are quenched {\it i.e.,} fixed according to some pattern  that may correspond to a macroscopic mass density profile $\varphi_q(\bar{x})$ in the thermodynamic limit. On the other hand, the velocities of the particles are chosen from some distribution $h_q(u|x)$ [see Eq.~\eqref{en-qa}]. To proceed, we perform average of $\mathscr{P}_i(z,v,t|\bar{{\bf x}},{\bf u},0)$, given in Eq.~\eqref{def:q_p-q-m}, over the initial velocities $\{u_{i}\}$ and get  
 \begin{align}
 \begin{split}
 \mathtt{P}_i(z,v,t|\bar{{\bf x}}) &= \prod_{k=1}^N \int du_k \mathscr{P}_i(z,v,t| \bar{{\bf x}},{\bf u},0) h(u|\bar{x}_k), \\
 &=\sum_{m=1}^N \sum_{S_\ell, S_r}  {g}(z,v,t|\bar{x}_m,0)~\prod_{k=1}^{i-1} {g}_{<}(z,t|\bar{x}_{S_\ell[k]},0) 
\prod_{k=i+1}^{N} {g}_{>}(z,t|\bar{x}_{S_r[k]},0),  
\end{split}
\label{mttP_q}
\end{align}
where
\begin{subequations}
\begin{align}
{g}(y,v,t|\bar{x},0)&= \int du \ g(y,v,t|\bar{x},u,0) { h_q(u)}, \\
~~ {g}_{<}(z,t|\bar{x},0) &= \int_{-\infty}^z dy\int_{-\infty}^\infty dv~{g}(y,v,t|\bar{x},0), 
\end{align}
\text{and}
\begin{align}
{g}_{>}(z,t|\bar{x},0) &=1-{g}_{<}(z,t|\bar{x},0). 
 \end{align}
\label{def:barg_{<}-barg_{>}}
\end{subequations}
We introduce variables $\epsilon_k$ and functions ${\mathtt{g}}_{\epsilon_k}$ for each $k$ such that, one has  ${\mathtt{g}}_0={g}_<$ and ${\mathtt{g}}_1={g}_>$ for the two values $\epsilon_k=0$ and $\epsilon_k=1$, respectively. 
In terms of these functions, we can rewrite  Eq.~\eqref{mttP_q} as 
\begin{align}
\mathscr{P}_i(z,v,t|\bar{{\bf x}},0) 
&=\sum_{m=1}^N  {g}(z,v,t|\bar{x}_m)~\prod_{\substack{k=1 \\ k\ne m}}^{N}  \left( \sum_{\epsilon_k=0,1} \right) \delta_{\sum_{k=1}^N\epsilon_k,(N-i)}~ {\mathtt{g}}_{\epsilon_k}(z,t|\bar{x}_{k},0), \notag \\ 
&=\frac{1}{2\pi} \int_{-\pi}^\pi d\vartheta~e^{-\iota(i-1)\vartheta}
\sum_{m=1}^N {g}(z,v,t|\bar{x}_m,0)  \prod_{\substack{k=1\\k\ne m}}^N \left(1+(e^{\iota \vartheta}-1) {g}_{<}(z,t|\bar{x}_k,0)\right), \notag 
\end{align}
In the third line of the above equation we have used the integral representation of Kronecker  delta $\delta_{n,0} = \left(1/2 \pi \right) \int_{-\pi}^\pi d \vartheta \ e^{\iota n \vartheta}$ with $\iota^2=-1$ and ${g}_{<} =1 -{g}_{>}$.

In the thermodynamic limit ($N \to \infty$), one can simplify the above expression further. First we write the product over $k$ inside the integral as exponential of a sum over $\bar{x}_k$. 
{In the limit of large $N$, one can approximate such sums by integrals over $\bar{x}$ as $\sum_k\mathfrak{f}(\bar{x}_k) \approx \int d\bar{x}~\varphi_q(\bar{x})~\mathfrak{f}(\bar{x})$ where $\varphi_q(\bar{x}) \simeq N~p_q(\bar{x})$ with $p_q(\bar{x})=\lim_{N\to \infty}\frac{1}{N}\sum_k\delta(\bar{x}-\bar{x}_k)$ [also see after Eq.~\eqref{en-qa}].  We finally get }
\begin{align}
 \mathtt{P}_i(z,v,t) & =  \int_{-\pi}^\pi \frac{d\vartheta}{2\pi}e^{-\iota(i-1)\vartheta} \exp \left[ \int d\bar{x} \varphi_q(\bar{x}) \ln \left(1+(e^{\iota \vartheta}-1) {g}_{<}(z,t|\bar{x},0)\right)\right] ~f^0(z,V,t) \nonumber 
 \end{align}
 with $f^0(z,V,t) =  \int d\bar{x} \ \varphi_q(\bar{x})~{g}(z,V,t|{\bar{x}},0)$,
where $\varphi_q(\bar{x})$ is the mass density in which the particles were initially arranged in the quenched configuration. Using this form of 
$ \mathtt{P}_i(z,v,t) $ in  Eq.~\eqref{def:x*_i} and {again approximating the sum over $i$ by an} integral, we find 
\begin{align}
\begin{split}
f(X,V,t) &= \int_{-\infty}^\infty \frac{dz}{2\pi} \left[  \int_{-\infty}^\infty ~d \alpha ~e^{-\iota(X-z) \alpha} \right. \\ 
&~~~\times~\left. \exp \left( \int d\bar{x} \varphi_q(\bar{x}) \ln [1+(e^{\iota \alpha a}-1){g}_{<}(z,t|\bar{x},0)]\right) \right]~f^0(z,V,t),
\end{split}
\end{align}
where, recall $\varphi_q(\bar{x})\simeq N p_q(\bar{x})$.
Expanding the exponent $e^{\iota \alpha a}$ to linear order in $\alpha a$ and performing the integral over $\alpha$, we get the Euler solution as in Eq.~\eqref{sol:euler-micro} with $F^0(z,t)=\int d\bar{x}\varphi_q(\bar{x}){g}_{<}(z,t|\bar{x},0)$ where ${g}_{<}(z,t|\bar{x},0)$ is defined in Eq.~\eqref{def:barg_{<}-barg_{>}}. 
{\color{black} To get the finite $N$ correction,  as done in the annealed case, we expand the exponent to quadratic order in $\alpha a$ and then take the large $N$ limit keeping $Na$ fixed.  We find the following $O(1/N)$ correction 
\begin{align}
f(X,V,t) & \approx \int_{-\infty}^\infty \frac{dz}{2\pi} \left[  \int_{-\infty}^\infty ~d \alpha ~e^{-\iota(X-z) \alpha} \exp \left(  \iota \alpha a F^0(z,t) - \frac{\alpha^2a^2}{2N} \Sigma_q^2(z,t)\right) \right]~f^0(z,V,t),  \label{f(x,v,t)-quenched-1/N}
\end{align}
with
\begin{align}
\begin{split}
\Sigma_q^2(z,t)&=N \int d\bar{x} \varphi_q(\bar{x}) \left({g}_{<}(z,t|\bar{x},0)-{g}_{<}^2(z,t|\bar{x},0) \right), \\ 
&= 
NF^0(z,t) - N\int d\bar{x} ~\varphi_q(\bar{x})~ {g}_{<}^2(z,t|\bar{x},0).
\end{split}
\label{Sigma_q^2}
\end{align} 
Performing the integral over $\alpha$ in Eq.~\eqref{f(x,v,t)-quenched-1/N},  once again we find an expression for $f(X,V,t)$ in which a Gaussian substructure replaces the delta function in the Euler solution in Eq.~\eqref{sol:euler-micro} to convolute with the point particle distribution function$f^0(z,V,t)$ as in Eq.~\eqref{sol:beyond-Euler-2}. However, in the quenched case one gets a Gaussian with different variance:
\begin{align}
{f(X,V,t) 
\approx \int_{-\infty}^\infty dz \frac{\sqrt{N}\exp \left(-\frac{N(X-z-a F^0(z,t))^2}{2a^2\Sigma_q^2(z,t)} \right)}{\sqrt{2 \pi a^2\Sigma_q^2(z,t)}}~f^0(z,V,t).}
\label{sol:beyond-Euler-q}
\end{align}
It turns out that the variance $\Sigma_q^2(z,t)$ in Eq.~\eqref{Sigma_q^2} is smaller than the variance $\Sigma_a^2(z,t)$ for the annealed case  in Eq.~\eqref{Sigma_a^2}. To see this one needs to compare the variances for the annealed and quenched initial conditions  corresponding to the same mass density profile $\varphi_q(x)=\varphi(x)=Np(x)$. 
Comparing $\Sigma_a^2(z,t)$ for annealed  initial configurations with velocities chosen independently of the position with $\varphi_a(x)=\varphi(x)$, it is easy to show  that  $\Sigma_a^2(z,t) \ge \Sigma_q^2(z,t)$. This is because $\left( \int dx~p(x)g_<(z,t|x,0)\right)^2 \leq \int dx ~p(x)~g_<^2(z,t|x,0)$ for $0\leq p(x)<1$ and $0 \leq g_<(z,t|x,0)<1$ with $\int dx~p(x)=1$. Clearly the variance of the distribution $f(X,V,t)$ for $X$ is smaller in the quenched case than the annealed case (given in Eq.~\eqref{sol:beyond-Euler-2}). This difference can be seen more explicitly in the tracer particle problem discussed in the next section. }

\begin{figure}[h]
    \centering
    \includegraphics[scale = 0.65]{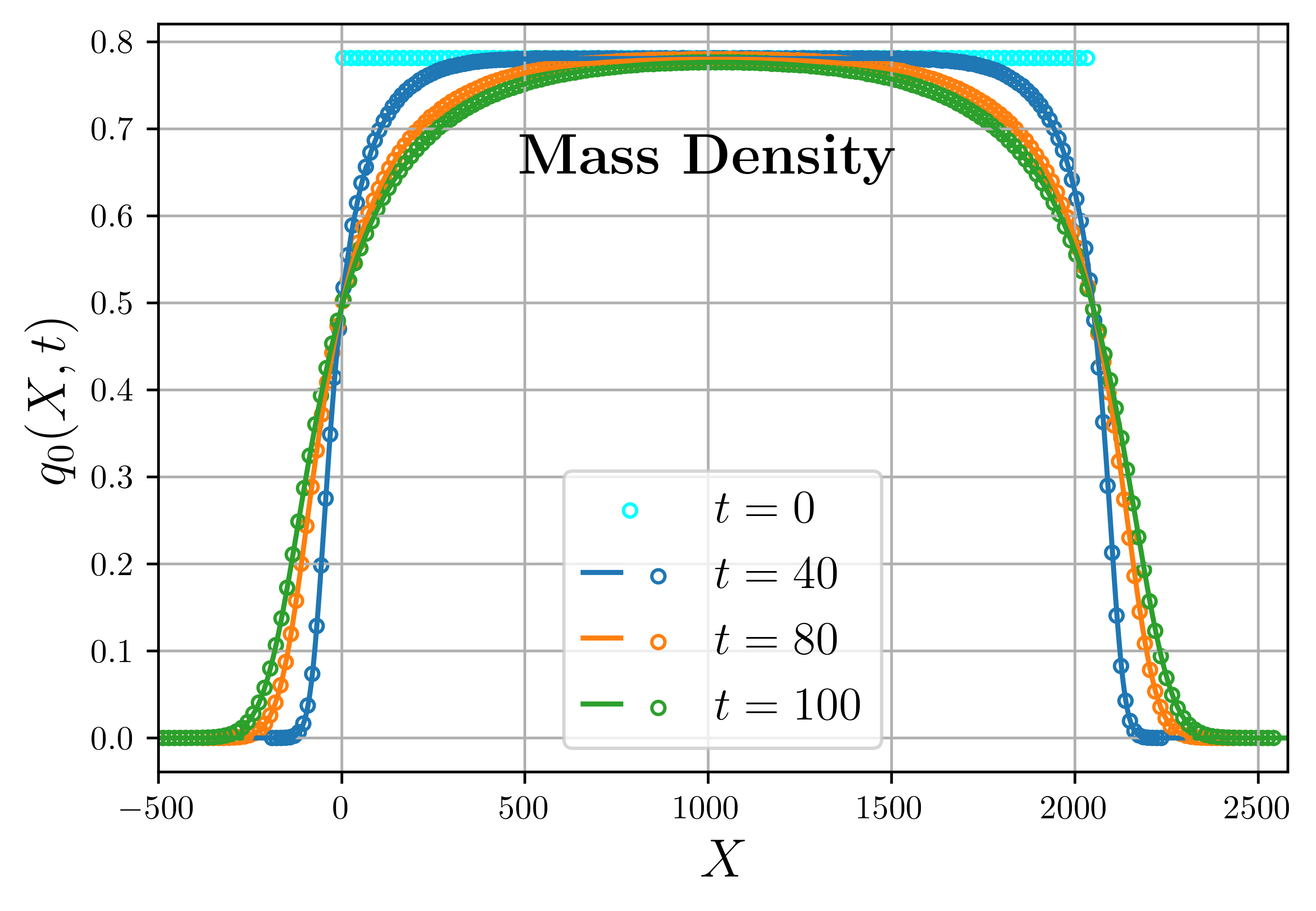}
    \caption{Plot showing the density profile of $N = 1600$ hard rods initially distributed uniformly in position in the range $[0,2049]$ and velocity chosen from Gaussian distribution in Eq.~\eqref{eq:max-dist} with $T=1.0$. The circles denote results from numerical simulation obtained after averaging over $10^5$ realizations where the initial distribution in position is quenched. The solid line denote results from the expression \eqref{sol:beyond-Euler-q}. We observe that they are in excellent agreement. }
    \label{fig:den-prof-quenched}
\end{figure}

In Fig.~\ref{fig:den-prof-quenched} we compare our analytical expression in Eq.~\eqref{sol:beyond-Euler-q} with numerical simulation for a quenched configuration of the positions corresponding to a uniform mass density profile {and velocities chosen from Maxwell distribution in Eq.~\eqref{eq:max-dist}.} We observe good agreement.

\section{Dynamics of a tracer rod}
\label{sec:tracer-dyna}
In this section we discuss dynamics of a single tracer rod moving in the background of many hard rods. This problem was 
first studied by Lebowitz, Percus and Sykes \cite{lebowitz1968time}. They considered a gas of hard rods in equilibrium with the restriction that there was a tracer particle with velocity $v_0$ at the origin at $t=0$. We emphasize that at an arbitrary time, the tracer rod is the rod with velocity $v_0$ which could be completely different from the initial  rod. This is different from a tagged particle in single file motion where the tagged particle does not lose its initial label as was assigned initially according to their positions \cite{hegde2014universal, krapivsky2015tagged, cividini2017tagged}. In our case, the tag of the rod is its velocity $v_0$.

In due course of time, this tracer rod will move ballistically in between collisions with other rods. At each collision, the tracer tag of the rod jumps to the other rod with which it exchanges velocity and consequently, its position jumps by the rod length. It is important to recall that such jumps do not occur for point particles   {\it i.e.,} when rod  length  is zero and the displacement of the tracer particle is always $v_0t$ irrespective of the configuration of other particles. On the other hand for hard rods, the amount of extra displacement made by the tracer rod (rod with velocity $v_0$) in addition to the ballistic drift,  is non zero and it fluctuates from one configuration of the background rods to other. These fluctuations give rise to the spreading of the distribution of the position of the tracer rod with time and at the hydrodynamic level such spreading is supposed to be captured by the NS correction to the Euler GHD as in Eq.~\eqref{eq:GHD-NS}. LPS had shown that the distribution of the position of tracer rod is a Gaussian at late times and the diffusion constant depends on the velocity distribution of the background particles \cite{lebowitz1968time}.  The same result has recently been obtained by solving the NS equation \eqref{eq:GHD-NS} starting from the same initial condition as LPS \cite{singh2024thermalization}. In this paper we re-derive this result using an alternate microscopic approach. While the derivation in \cite{lebowitz1968time} and \cite{singh2024thermalization}, rely on equilibrium of  background rods characterized by homogeneous position distribution,  we generalise our derivation to initial conditions, both annealed and quenched, with inhomogeneous mass density profile.

We start by noting that for any given initial configuration of the background rods, the displacement of the tracer rod at time $t$, starting from the origin, is 
\begin{align}
X(t) = v_0t + a[\mathcal{N}_{r\ell}(t) - \mathcal{N}_{\ell r}(t)], \label{X_t-tracer}
\end{align}
where $\mathcal{N}_{r\ell}(t)$ is the number of rods that collided with the tracer rod from the right and similarly,  $\mathcal{N}_{\ell r}(t)$ is the number of rods that collided from the left of the tracer rod. At each collision from the right, the tracer rod jumps to the right by $a$. On the other hand, at each collision from the left, the tracer rod jumps by amount $a$ to the left and this is the reason why there is a minus sign in front of third term in Eq.~\eqref{X_t-tracer}. We aim to find the distribution of $X$, denoted  by $\mathds{P}_t(X)$ at time $t$. To find this distribution,  we need to find the joint distribution of $\mathcal{N}_{r\ell}(t)$ and $\mathcal{N}_{\ell r}(t)$. At this point, mapping the dynamics of hard rods to point particles, as done previously, becomes useful. This is because, in the point particle representation, the tracer point particle will encounter exactly the same number of collisions ($\mathcal{N}_{r\ell}(t)$ from right and $\mathcal{N}_{\ell r}(t)$ from left) 
{till time $t$} as that of the tracer rod in the corresponding hard rod configuration. Hence, we compute the joint distribution of $\mathcal{N}_{r\ell}(t)=n$ and $\mathcal{N}_{\ell r}(t)=m$, which we denote by $\mathcal{P}(n,m,t)$, in the point particle picture. For this we follow a procedure used in the computation of tagged particle motion in single file problem \cite{krapivsky2015tagged, cividini2017tagged}.

We assume that initially there are $N$ point particles on the left  and $N$ point particles on the right of the tracer which is located at the origin on a {one-dimensional} straight line. Let the positions of these point particles be denoted by $\{\bar{x}_{k}\}$ with $k = -N,...,-1,1,...,N$ such that {$\bar{x}_k<0$} if $k<0$ and $\bar{x}_k>0$ if $k>0$ . The velocities of the particles are chosen independently from some velocity distribution $h(v)$ except for the tracer particle which has velocity $v_0$. At time $t$ this particle will reach position $v_0t$. Other particles will reach random positions as their velocities are random. Let $\mathfrak{g}(y,t|\bar{x},0)$ represent the propagator for a point particle to reach $y$ at time $t$ starting from $\bar{x}$. It is easy to see that the propagator can be written as 
\begin{align}
\mathfrak{g}(y,t|\bar{x},0) =\frac{1}{\varsigma_t} h\left( \frac{y-\bar{x}}{\varsigma_t}\right),  \label{propagator}
\end{align}
where for a ballistic particle $\varsigma_t=t$. Note  that 
\begin{align} 
{g}_{<}(z,t|\bar{x},0) = \int_{-\infty}^z dy~\mathfrak{g}(y,t|\bar{x},0), \label{def:barg_{<}}
\end{align}
 represents the probability that a particle, starting from $\bar{x}$, can be found below $z$ at time $t$. Similarly, 
 \begin{align} 
{g}_{>}(z,t|\bar{x},0) = \int_z^{\infty} dy~\mathfrak{g}(y,t|\bar{x},0), \label{def:barg_{>}}
\end{align}
 is the probability of finding it above $z$ at time $t$. Note these are essentially same quantities as defined in Eq.~\eqref{def:barg_{<}-barg_{>}} and hence bear the same notation.
In terms of these probabilities, one can now write the joint probability $\mathcal{P}(n,m,t)$ as 
\begin{align}
\begin{split}
\mathcal{P}(\mathcal{N}_{r\ell}(t)=n,\mathcal{N}_{\ell r}(t)=m,t) &=
 \left[ \prod_{j=1}^{N} \left(\sum_{\epsilon_{j}=1,0}\right) \delta_{\sum_{j=1}^{N}\epsilon_{j},(N-n)} ~{\mathtt{g}}_{\epsilon_{j}}(v_0t,t|\bar{x}_{j},0) \right] 
\\&~~ \times
~ \left[ \prod_{k=1}^{N} \left(\sum_{\epsilon_{-k}=1,0}\right) \delta_{\sum_{k=1}^{N}\epsilon_{-k},m} ~{\mathtt{g}}_{\epsilon_{-k}}(v_0t,t|\bar{x}_{-k},0) \right], 
\end{split}
\label{def:mcalP(n,m,t)-0}
\end{align}
where ${\mathtt{g}}_{\epsilon}=\delta_{\epsilon,0}{g}_< + \delta_{\epsilon,1}{g}_>$.
The Kronecker delta in the first line makes sure that $N-n$ number of particles, starting above origin have reached above position $v_0t$ at time $t$. Similarly, the Kronecker delta in the second line makes sure that $m$ number of particles, starting below origin have reached above position $v_0t$ at time $t$.
Once again introducing integral representation  $\delta_{n,0} = \left(1/2 \pi \right) \int_{-\pi}^\pi d \vartheta e^{\iota n \vartheta}$ of Kronecker delta and performing some simplifications, we can rewrite Eq.~\eqref{def:mcalP(n,m,t)-0} as 
{
\begin{align}
\mathcal{P}(n,m,t) &= \mathcal{P}_r(n,t)~\mathcal{P}_\ell(m,t),  \label{def:mcalP(n,m,t)}
\end{align}
where
\begin{align}
\begin{split}
 \mathcal{P}_r(n,t)&=\frac{1}{2 \pi }\int_{-\pi}^\pi d\vartheta ~e^{-\iota n \vartheta}\prod_{k=1}^{N}\left[1+(e^{\iota \vartheta}-1){g}_{<}(v_0t,t|\bar{x}_{k},0) \right] \\
\mathcal{P}_\ell(m,t)&=\frac{1}{2 \pi } \int_{-\pi}^\pi d\xi  ~e^{-\iota m \xi}\prod_{k=1}^{N}\left[1+(e^{\iota \xi}-1){g}_{>}(v_0t,t|\bar{x}_{-k},0) \right] .
\end{split}
\label{def:mcalP_lr(n,m,t)}
\end{align} }
\noindent Once $\mathcal{P}(n,m,t)$ is known, the moment generating function (MGF) of the displacement $X(t)$ of the tracer particle given in Eq.~\eqref{X_t-tracer}, can be computed as 
{
\begin{align}
\mathfrak{Z}(\alpha,t) &= \left \langle e^{-\iota \alpha X} \right \rangle
=e^{-\iota \alpha v_0 t}~ \mathfrak{Z}_r(\alpha,t)~\mathfrak{Z}_\ell(-\alpha,t) \label{def:charac-func}  
\end{align}
with $\mathfrak{Z}_{r,\ell}(\alpha,t)=\sum_{n=0}^Ne^{-\iota a\alpha n}~\mathcal{P}_{r,\ell}(n,t)$. }
To proceed from here we need to discuss the annealed and quenched initial conditions separately.

\subsection{Annealed initial condition}
\label{sec:tracer-anld-ic}
\begin{figure}[h]
    \centering
    \includegraphics[width=0.98\textwidth]{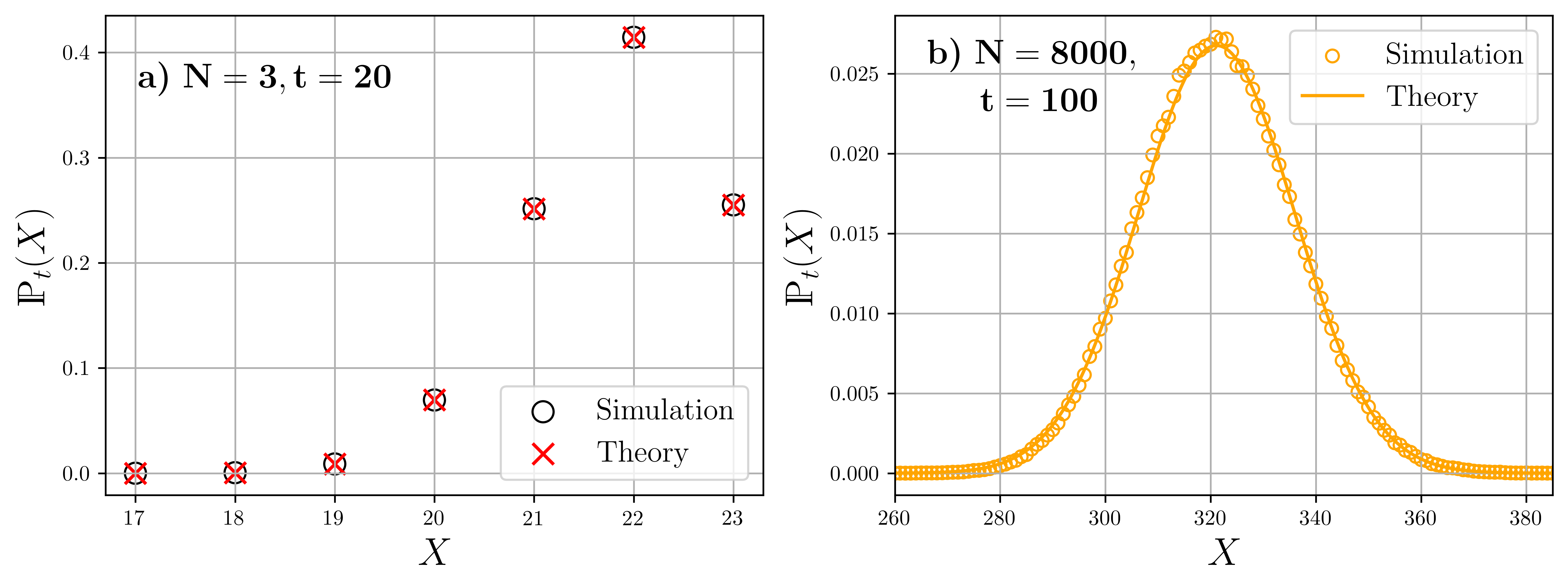}
    \caption{Plots showing comparison between analytical result in Eq.~\eqref{mfrkP-exact} for the probability $\mathds{P}_t(X)$ of finding the tracer rod at position $X$ at time $t$, with the same obtained from numerical simulation for annealed initial condition. The plot (a) corresponds to $N=3$ and $t=20$ whereas plot (b) corresponds to $N=8000$ and $t=100$. The initial individual positions of the background rods are chosen independently in the PP picture according to Eq. \eqref{eq:initialgaussianpos}. In plot (a), the parameters are $l_0 = 10$ and $\sigma = 5$ while in plot (b), the parameters are $l_0 = 300$ and $\sigma = 30$.  The simulation data  are obtained after averaging over $10^7$ initial configurations in (a) and over $5 \times 10^5$ realizations in (b). For both the plots, other parameters used are  $a = 1.0$ and $v_0 = 1$, and the velocities of background rods are chosen from Maxwell distribution in Eq.~\eqref{eq:max-dist} with $T=1.0$.}
    \label{fig:Quenched_gaussian}
\end{figure}

In this case, the initial positions of the particles are chosen randomly from some distribution. We first put the tracer particle at the origin. Then we draw positions of $N$ particles on the left of the tracer particle randomly and independently from a distribution $\frac{\varphi_\ell(\bar{x})}{N}$ such that the average initial density on the left is $\varphi_\ell(\bar{x})$. Similarly, the positions of the particles on the right are chosen randomly and independently from the distribution $\frac{\varphi_r(\bar{x})}{N}$ so that the density on the right is $\varphi_r(\bar{x})$. Performing average over these initial distributions one can simplify the expression in Eq.~\eqref{def:mcalP(n,m,t)} considerably. We get,
{
\begin{flalign}
\mathcal{P}(n,m,t) &= \binom{N}{n}~ \left[\frac{p_{r\ell}}{N}\right]^n\left[\frac{p_{rr}}{N}\right]^{(N-n)}\times~\binom{N}{m}~ \left[\frac{p_{\ell r}}{N}\right]^m\left[\frac{p_{\ell \ell}}{N}\right]^{(N-m)}, 
\label{mcalP(n,m,t)-an} 
\end{flalign}
where
\begin{align}
\begin{split}
p_{r\ell}(t) &= \int_0^\infty d\bar{x}~{g}_{<}(v_0t,t|\bar{x},0)~\varphi_r(\bar{x}),\\ 
p_{rr}(t) &= \int_0^\infty d\bar{x}~{g}_{>}(v_0t,t|\bar{x},0)~\varphi_r(\bar{x}) =N - p_{r\ell}, \\
p_{\ell \ell}(t) &= \int_{-\infty}^0 d\bar{x}~{g}_{<}(v_0t,t|\bar{x},0)~\varphi_\ell(\bar{x}),\\ 
p_{\ell r}(t) &= \int_{-\infty}^0 d\bar{x}~{g}_{>}(v_0t,t|\bar{x},0)~\varphi_\ell(\bar{x}) =N - p_{\ell \ell}.
\end{split}
\label{def:part-probs} 
\end{align} }
This product of two binomial distributions can be easily understood from the fact that the dynamics of hard-point particles can further be mapped to non-interacting point particles via Jepsen mapping \cite{jepsen1965dynamics}.
Inserting the above form of $\mathcal{P}(n,m,t)$ from Eq.~\eqref{mcalP(n,m,t)-an} in Eq.~\eqref{def:charac-func}, and performing the summations we get 
\begin{align}
{\mathfrak{Z}(\alpha,t) = e^{-\iota \alpha v_0 t} \bigg[1+(e^{-\iota \alpha a}-1) \frac{p_{r \ell}}{N} \bigg]^N~\times~\bigg[1+(e^{\iota \alpha a}-1) \frac{p_{\ell r}}{N}\bigg]^N.}
\end{align}
Note that, by definition one should be able to expand the function  {$e^{\iota \alpha v_0 t}\mathfrak{Z}(\alpha ,t)$ in integer powers of $e^{-\iota \alpha}$.  From the coefficient of $e^{-\iota \alpha Y}$} where $Y=X-v_0t$, one can find the probability of $X$ as 
\begin{align}
\mathds{P}_t(X=Y+v_0t) = \sum_{n=0}^N \binom{N}{\frac{Y + na}{a}}  \binom{N}{n} \bigg[ \frac{p_{r r}}{N} \bigg]^{N - \frac{Y + na}{a}}~ \bigg[ \frac{p_{r \ell}}{N} \bigg]^{\frac{Y + na}{a}}~\bigg[\frac{p_{\ell \ell}}{N} \bigg]^{N - n}~\bigg[\frac{p_{\ell r}}{N} \bigg]^n,
\label{mfrkP-exact}
\end{align}
where $-N \le Y \le N$.
This is an exact microscopic expression for the distribution $\mathds{P}_t(Z)$ valid for arbitrary number of particles $N$ and arbitrary choices of initial particle density profiles $\varphi_{\ell }(\bar{x})$ and $\varphi_{r}(\bar{x})$. In Fig.~\ref{fig:Quenched_gaussian}a we provide a numerical verification of the above expression for $2N+1=7$ rods. Initially, we place three rods each on both sides of the tracer rod at the origin. The positions of the rods are determined  from the corresponding point particle configurations. The  positions of the point particles are chosen independently from the following distribution
\begin{align}\label{eq:initialgaussianpos}
p_a(x)=\frac{1}{\mathscr{Z}(l_0)}\left(\Theta(x)\frac{e^{-\frac{(x-l_0)^2}{2\sigma^2}}}{\sqrt{2 \pi \sigma^2}} + \Theta(-x)\frac{e^{-\frac{(x+l_0)^2}{2\sigma^2}}}{\sqrt{2 \pi \sigma^2}}\right),
\end{align}
where $\mathscr{Z}(l_0)$ is the normalisation with $l_0=10$ and $\sigma=5$ and then converted to hard rod positions following the procedure described above and around Eq.~\eqref{inv-map-to-hpg-line}. The velocities of these particles are chosen from Maxwell distribution, $h(v)$ given in Eq.~\eqref{eq:max-dist} with $T=1.0$. The excellent agreement verifies our analytical result.

While the result in Eq.~\eqref{mfrkP-exact} is exact and explicit, in order to find the important features of the motion of the tracer particle in the hydrodynamic limit we need to consider  {large $N$ and small $a$ limit keeping $Na$ fixed.} In this limit, the expression of the MGF simplifies to 
\begin{align}
\mathfrak{Z}(\alpha,t)=\left \langle e^{-\iota \alpha X} \right \rangle = \exp \left[ -\iota \alpha v_0t +(e^{-\iota \alpha a}-1)~p_{r \ell}(t)+ (e^{\iota \alpha a}-1)~p_{\ell r}(t) \right].
\end{align}
Now, as done before expanding the $e^{\pm \iota \alpha a}$ factors to quadratic order we find 
\begin{align}
\mathfrak{Z}(\alpha,t) \approx \exp\left( -\iota \alpha \left[v_0t +a p_{r\ell }(t)-ap_{ \ell r}(t)\right] -\frac{1}{2}a^2 \left[p_{\ell r}(t)+p_{r \ell}(t)\right]~ \alpha^2 \right)
\end{align}
which can be inverted to provide a gaussian form for the distribution $\mathds{P}_t(Z)$ as 
\begin{align}
\mathds{P}_t(X) &= \frac{1}{\sqrt{2 \pi \Sigma_a^2(t)}}~\exp\left( - \frac{(X - \langle X \rangle)^2}{2 \Sigma_a^2(t)}\right),~~\text{where}  \label{P(Z)-gauss}\\
\langle X \rangle &= v_0t +a [p_{r \ell}(t)-p_{\ell r}(t)],~\text{and}~\Sigma_a^2(t) = a^2 [p_{\ell r}(t)+p_{r \ell}(t)], \label{mean-var-Z}
\end{align}
with $p_{r\ell}(t)$ and $p_{\ell r}(t)$ being provided in Eq.~\eqref{def:part-probs}. Note, from the above expression one can reproduce the distribution of the displacements of tracer rod obtained in \cite{lebowitz1968time, singh2024thermalization} for specific choice of uniform density of the background rods, say $\varrho_0$ which gets transformed to uniform density $\varphi_0 =\frac{\varrho_0}{1-a\varrho_0}$ for point particles. For such initial state, the expressions of $p_{r\ell}(t)$ and $p_{\ell r}(t)$ in Eq.~\eqref{def:part-probs} get simplified and one has 
\begin{align}
p_{r\ell}&=\frac{\varrho_0}{1-a \varrho_0} t \int_{-\infty}^{v_0} dv~(v_0-v)h(v), \\
p_{\ell r}&= \frac{\varrho_0}{1-a \varrho_0}  t \int_{v_0}^{\infty} dv~(v-v_0)h(v).
\end{align}
Inserting these expressions in Eq.~\eqref{mean-var-Z}, one gets 
\begin{align}
\langle X\rangle &= \frac{v_0-a \varrho_0 u}{1-a \varrho_0}~t,~~\text{with}~u=\int_{-\infty}^\infty dv~v~h(v), \label{<Z>}\\
\Sigma_a^2(t) &=\frac{\varrho_0~t}{1-a\varrho_0}~\int_{-\infty}^\infty dv~|v-v_0|~h(v). \label{Sigma-a^2}
\end{align}
Note that the mean velocity $\langle X \rangle /t$ extracted from Eq.~\eqref{<Z>} is essentially equal to  $V_{\rm eff}$ as given in Eq.~\eqref{def:V_eff}. Also note that the diffusion constant $D_a=\Sigma^2/t$ extracted from  Eq.~\eqref{Sigma-a^2} agrees with the expression obtained in \cite{singh2024thermalization} by solving the Navier-Stokes equation \eqref{eq:GHD-NS}. The results in Eqs.~\eqref{P(Z)-gauss} and \eqref{mean-var-Z} generalises these results  for arbitrary initial density profiles $\varphi_\ell(\bar{x})$ and $\varphi_r(\bar{x})$ and velocity distribution $h(v)$. 

\subsection{Position quenched initial condition}
In this case one does not choose the initial positions $\{\bar{x}_k\}$ from some distribution, instead they are arranged in such a way that in the thermodynamic limit they correspond to a macroscopic mass density profile $\varphi_q(\bar{x})$.  In general, the mass distribution function could be different on the $+$ve and $-$ve axes and we denote them by $\varphi_{q,r}(\bar{x})$ and $\varphi_{q,\ell}(\bar{x})$ respectively.  As done previously, we write the distributions in Eq.~\eqref{def:mcalP_lr(n,m,t)}  as  integral over  $\varphi_{q,r/\ell}(\bar{x})$  in the $N \to \infty$ limit. This can be done by writing the products over $k$ in Eq.~\eqref{def:mcalP_lr(n,m,t)} as exponential of sum over $k$ and then {approximating this sum by }an integral over $\bar{x}$ with density $\varphi_{q,r/\ell}(\bar{x})$. In the large $t$ limit, it is easy to realise that the distributions $\mathcal{P}_{q,r/\ell}(n,t)$ would be highly peaked around large values of $n$. This limit can be obtained equivalently by taking small $a$ limit, in which case Eq.~\eqref{def:mcalP_lr(n,m,t)} can be written as 
\begin{align}
\begin{split}
 \mathcal{P}_r(n,t)&=\int_{-\infty}^\infty \frac{d\alpha}{2\pi} ~e^{-\iota n \alpha} \exp\left(\int_0^\infty d\bar{x} ~\varphi_{q,r}(\bar{x}) \ln\left[1+(e^{\iota  \alpha a}-1){g}_{<}(v_0t,t|\bar{x},0) \right] \right)\\
\mathcal{P}_\ell(m,t)&= \int_{-\infty}^\infty \frac{d\alpha}{2\pi} ~e^{-\iota m \alpha}~\exp\left(\int_{-\infty}^0 d\bar{x} ~\varphi_{q,\ell}(\bar{x}) \ln\left[1+(e^{\iota  \alpha a}-1){g}_{>}(v_0t,t|\bar{x},0) \right] \right).
\end{split}
\label{mcalP_lr-2}
\end{align}
\begin{figure}[t]
    \centering
    \begin{subfigure}{0.48\textwidth}
        \centering
        \includegraphics[width=\textwidth]{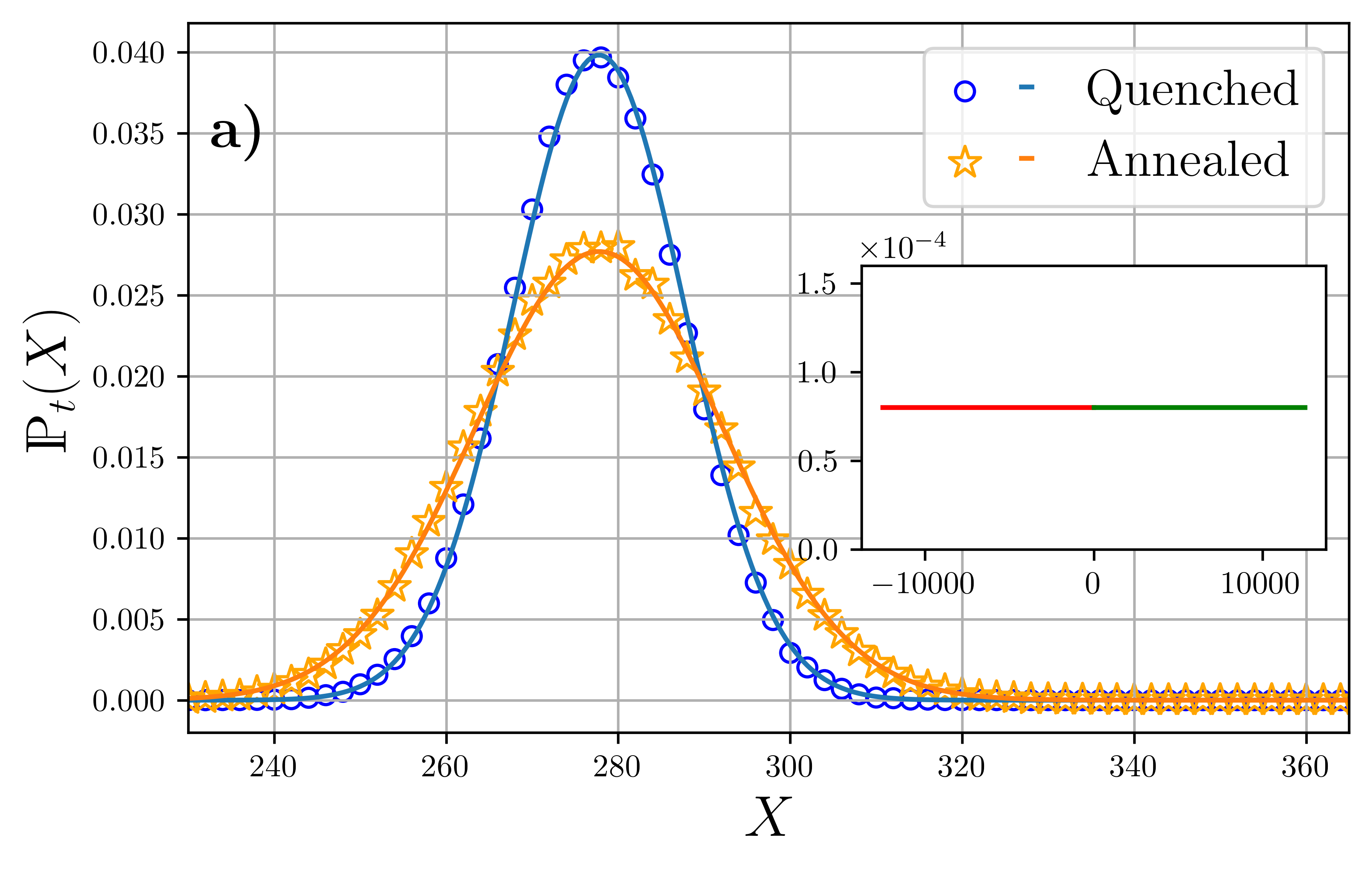}
    \end{subfigure}
    \hfill
    \begin{subfigure}{0.48\textwidth}
        \centering
        \includegraphics[width=\textwidth]{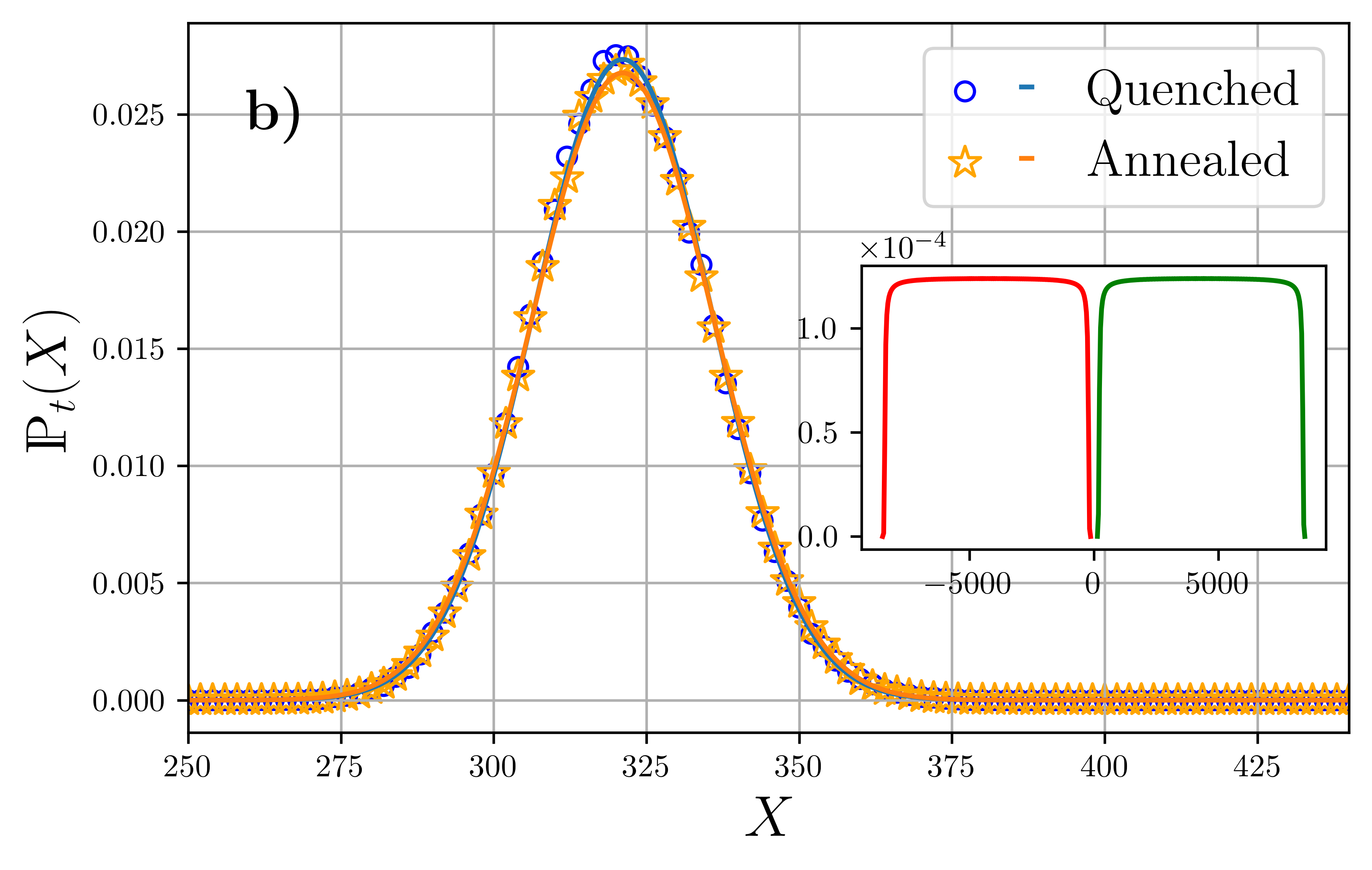}
    \end{subfigure}
    \caption{Plots showing comparison between simulation (circles)  and  theoretical result from Eq.~\eqref{P-q(Z)-gauss} (solid lines) for the probability $\mathds{P}_t(X)$ for two quenched initial conditions. The plot (a) corresponds to uniform mass density on both sides of the tracer at the origin  initially (see inset) whereas plot (b) corresponds to doubly peaked mass density profile at $t=0$. The two peaks are on the two sides of the tracer at the origin (see inset). In the PP picture, these two double peaks correspond to two Gaussian peaks on the two sides, respectively at $x=\pm 300$ with variance $900$. For comparison, we also plot  theoretical (from Eq.~\eqref{P(Z)-gauss}, orange dashed lines) and simulation data (stars) for annealed initial configuration corresponding to same mass density profiles. In both the plots (a) and (b), we have chosen $N = 8000$, $t=100$, $a = 1.0$ and $v_0 = 1$. Simulation data in both plots are averaged over $5\times10^5$ realizations.  We observe that the uniform distribution of hard rods on both sides display a significant difference between the annealed and quenched cases, however the one with the Gaussian distribution does not display such significant difference.}
    \label{fig:Annealed_Quenched_LargeN}
\end{figure}
From these two expressions, one can easily find that 
\begin{align}
 \mathfrak{Z}_r(\alpha,t) &= \exp\left(\int_0^\infty d\bar{x} ~\varphi_r(\bar{x}) \ln\left[1+(e^{\iota  \alpha a}-1){g}_{<}(v_0t,t|\bar{x},0) \right] \right), \\
 \mathfrak{Z}_\ell(\alpha,t) &=  \exp\left(\int_{-\infty}^0 d\bar{x} ~\varphi_\ell(\bar{x}) \ln\left[1+(e^{\iota  \alpha a}-1){g}_{>}(v_0t,t|\bar{x},0) \right] \right).
\end{align}
inserting which in Eq.~\eqref{def:charac-func}, we get the MGF $\mathfrak{Z}(\alpha,t)$ of  the distribution of the displacement $X(t)$ of the tracer rod. As done in the annealed case, expanding the $e^{\pm \iota \alpha a}$ to quadratic orders in $\alpha a$ {and then taking the large $N$, small $a$ limits keeping $Na$ fixed,} we get 
\begin{align}
\mathfrak{Z}(\alpha,t) &\approx \exp\left( -\iota \alpha \left[v_0t +a p_{r \ell}(t)-ap_{\ell r}(t)\right] -\frac{a^2\alpha^2}{2} \left[p_{\ell r}(t)+p_{r \ell}(t) - \mathtt{q}_{\ell r}(t)-\mathtt{q}_{r \ell}(t) \right]  \right), \label{mfrkZ(alpha,t)-q}
\end{align}
where $p_{\ell r}(t)$ and $p_{r \ell}(t)$ are same functions defined in Eq.~\eqref{def:part-probs} except now the distribution functions $\varphi_{r/\ell}(\bar{x})$ are replaced by $\varphi_{q,r/\ell}(\bar{x})$, and 
\begin{align}
\begin{split}
\mathtt{q}_{r\ell}(t) &= \int_0^\infty d\bar{x}~\varphi_{q,r}(\bar{x})~{g}_{<}^2(v_0t,t|\bar{x},0),\\
\mathtt{q}_{\ell r}(t) &= \int_{-\infty}^0 d\bar{x}~\varphi_{q,\ell}(\bar{x})~{g}_{>}^2(v_0t,t|\bar{x},0).
\end{split}
\label{q_lr}
\end{align}
Performing inverse Fourier transform of $\mathfrak{Z}(\alpha,t)$ in Eq.~\eqref{mfrkZ(alpha,t)-q}, we find the following Gaussian form for the $\mathds{P}_t(X)$ in the quenched case as 
{ \label{P-q(Z)-gauss}
\begin{align}
\mathds{P}_t(X) &= \frac{1}{\sqrt{2 \pi \Sigma_q^2(t)}}~\exp\left( - \frac{(Z - \langle Z \rangle)^2}{2 \Sigma_q^2(t)}\right),
\end{align}
where
\begin{subequations}
\label{q-mean-var-Z}
\begin{align}
\langle X \rangle &= v_0t +a [p_{r \ell}(t)-p_{\ell r}(t)],
\end{align}
\text{and}
\begin{align}
\Sigma_q^2(t) &= a^2 [p_{\ell r}(t)+p_{r \ell}(t) - \mathtt{q}_{\ell r}(t)-\mathtt{q}_{r \ell}(t) ].
\end{align}
\end{subequations}}
Note that the  variance of the displacement is smaller in the quenched case than that in the annealed case [see Eq.~\eqref{mean-var-Z}], however the mean remains the same. Once again the expressions become more explicit for uniform mass density $\varrho_0$ of the background rods for which case the corresponding density for point particles is also uniform with value $\varphi_0 =\frac{\varrho_0}{1-a\varrho_0}$. For this case one gets the same mean  $\langle X \rangle$ as given in Eq.~\eqref{<Z>}. However, the variance now is modified to  
{
\begin{align}
\Sigma_q^2(t) &=\Sigma_a^2(t) - 2 \frac{\varrho_0 ~t}{1-a\varrho_0} \left[ \int_{v_0t}^\infty dz ~H_-^2(z) +  \int_\infty^{v_0t} dz~ H_+^2(z) \right],  \label{Sigma-q^2} 
\end{align}
where
\begin{align}
H_\pm(z) = \int_{-\infty}^\infty dz'~\theta\left(\pm(z-z')\right)~h(z'). \label{def:H_pm}
\end{align}
}
The result in Eq.~\eqref{P-q(Z)-gauss}, corresponds to the distribution of the tracer position in thermodynamic limit. This result is verified numerically in Fig.~\ref{fig:Annealed_Quenched_LargeN} for two  (quenched) initial conditions. The figure on the left (figure a) corresponds to uniform  particle density profile  on both sides of the tracer particle (see the inset). On the other hand, for the plot on the right (figure b), we consider a different initial condition with two peaks on the two sides of the tracer rod at the middle. For comparison, we also plot the distributions of the tracer position in the annealed case (from Eq.~\eqref{P(Z)-gauss}) corresponding to the same initial density profiles. We observe that the distribution in quenched case is narrower than the one in annealed case. However, this effect is more pronounced in case of uniform initial profiles (Fig.~\ref{fig:Annealed_Quenched_LargeN}a). {Such dependence of the mean squared displacement of tracer particle on quenched or annealed  initial conditions has been studied well in single file problems in other systems \cite{rajesh2001exact,Leibovich2013everlasting,krapivsky2015dynamical,sadhu2015large,santra2024tracer,banerjee2022role}.}

\section{Evolution from domain wall initial condition}
\label{sec:domain-wall} 
In this section, we consider domain wall initial condition in which $2N$ rods are distributed over the box $[-L,L]$. 
On the left half of the box $[-L,0]$, 
we distribute $N$ rods uniformly with density $\varrho_0 =N/L$ and the remaining $N$ rods are also distributed  uniformly with density $\varrho_0 =N/L$ on the right half of the box $[0,L]$. 
Eventually we would like to take the thermodynamic limits $N \to \infty$ and $L \to \infty$ keeping $\varrho_0$ fixed. 
Each rod on the left half starts with a fixed velocity $v_0$ whereas the velocities of the rods on the right half are chosen from a distribution $h(V)$. Note that this initial condition is slightly different from the one discussed in Eq.~\eqref{en-aa}. The initial phase space density is given by 
\begin{align}
\begin{split}
f(X,V,t=0) &= \varrho_0 \left[\Theta\left(-X\right) \Theta\left(X+L\right)~\delta(V-v_0) + \Theta\left(X\right) \Theta\left(L-X\right)~h(V)\right], \\
&=g(X,0)\delta(V-v_0) + f_b(X,V,0)
\end{split}
\label{ic:domain-wall}
\end{align} 
Note that the phase space density at $t=0$ has two parts $f(X,V,0) = f_s(X,V,0) +f_b(X,V,0)$. The  first part  $ f_s(X,V,0)$ corresponds to the phase space density for the special rods with velocity $v_0$ on the left half of the box {with mass density $g(X,0)=\varrho_0 \Theta\left(-X\right) \Theta\left(X+L\right)$.} The other part $f_b(X,V,0)$ corresponds to the rods {on the right half of the box with velocity chosen from distribution $h(V)$ and mass density $\rho_b(X,0)=\varrho_0  \Theta\left(X\right) \Theta\left(L-X\right)$.} We call these rods to be background rods (marked by the subscript `b'). These rods are then allowed to move on the infinite line. Since under the collision dynamics the velocities of the rods do not mix but only get exchanged, one would continue to observe two parts in the phase space distribution $f(X,V,t)$ at later time $t$. More precisely, we expect the following form for $f(X,V,t)$ 
\begin{align}
f(X,V,t) = g(X,t) \delta(V-v_0) + f_b(X,V,t),
\end{align}
at later time as well.
We are interested to find the mass density profiles of these two components denoted, respectively,  by $g(X,t)$ for the special rods and $\rho_b(X,t)$ for the background rods where 
\begin{align}
\rho_b(X,t)= \int_{-\infty}^\infty dV~f_b(X,V,t). \label{def:rho-b}
\end{align}
The density profiles  $g(X,t)$ and $\rho_b(X,t)$ of the special rods and the background rods, respectively, were determined recently in ~\cite{singh2024thermalization} by solving the Euler GHD equations. It was observed that the initial jump of the two density profiles at the interface remains there for all times in the Euler solutions. At the locations of these jumps, the Euler solutions disagree with the density profiles {obtained from numerical simulation} which show broadening  of the jumps. Here, using microscopic approach we compute the density profiles $g(X,t)$ and $\rho_b(X,t)$  that capture such broadening and thus go beyond the Euler solution. The microscopic approach we use in this case is also based on finding the joint distribution of the number of 
collisions with different types of rods as done in the previous section.

\subsection{Computation of density profile, $g(X,t)$ for special rods}

To compute $g(X,t)$, we look at the dynamics of a quasiparticle from the special component, starting from position $X_0 <0$  with velocity $V=v_0$.  Since all the special rods move with the same velocity $v_0$, they would not collide with each other. The special rods will, however, undergo collisions with the background rods approaching them from the right. At each collision, the {special} quasiparticle jumps by rod length $a$ to the right. Hence, at time $t$ the position of this quasiparticle is 
\begin{align}
X(t)= X_0+v_0t + a \mathcal{N}_t, \label{def:Z(t)}
\end{align}
where $\mathcal{N}_t$ is number of background rods it has collided with in duration $t$ and this accounts for an extra displacement due to the finite extent of the rods. 
The number of collisions $\mathcal{N}_t$ fluctuates from one initial configuration of the positions (of the background particles) to another. To find the distribution of $X(t)$ given in Eq.~\eqref{def:Z(t)}, we need to find the distribution of $\mathcal{N}_t$ and for that we consider an equivalent point particle picture as done before.

In the point particle picture we construct an equivalent problem  in which we have $2N$ point particles starting from initial positions inside a smaller box $[-L',L']$ with $L'=L(1-a\varrho_0)$ and equally divided between the two halves. The point particles on the left half of this box have velocity $v=v_0$ and the particles on the right half have velocities chosen from $h(v)$. 
A rod at $X_0 \in [-L,L]$ maps to a point particle at position $x_0 \in [-L',L']$. The mapping from $X_0$ to $x_0$  and the inverse mapping are given by 
\begin{align}
x_0&=X_0(1-a\varrho_0),~~
X_0=x_0(1+a \varphi_0),~~\text{with}~~\varphi_0=\frac{\varrho_0}{1-a\varrho_0}. \label{mapp-z-Z}
\end{align}
Hence, corresponding to every {special} {quasiparticle} with the dynamics given by Eq.~\eqref{def:Z(t)}, there is a special {point particle} which has a ballistic dynamics unhindered by the number of background point particles it passes through. This {point} {particle} starts from position $x_0=X_0(1-a\varrho_0)$ with a velocity $v_0$ and reaches position $x(t)=x_0+v_0t$ at time $t$ during which it encounters exactly $\mathcal{N}_t$ number of collisions with background point particles which is the same as that the corresponding quasiparticle {in the} hard rod {picture} would face.

To compute the probability $\mathcal{P}(n,t|x)=\text{Prob.}(\mathcal{N}_t=n)$, we first note that the number of collisions $\mathcal{N}_t$ till time $t$ is the same as the number of background particles, starting from positions on the right of the origin, reaching positions below $x= x_0+v_0t$ at time $t$. The probability for one background particle to reach position below $w$, starting above position $w_0$, can be written in terms of the single particle propagator $\mathfrak{g}(y,t|\bar{x},0)$ in Eq.~\eqref{propagator} and it is given by 
{ 
\begin{align}
\mathtt{p}_{r \ell} &= \frac{p_{r \ell}(w,t|w_0)}{N},
\end{align}
with
\begin{align}
p_{r \ell}(w,t|w_0)&= \int_{w_0}^\infty d\bar{x}~\varphi_0 \int_{-\infty}^wdy~\mathfrak{g}(y,t|\bar{x},0) = \varphi_0 ~t\int_{-\infty}^\psi dv(\psi-v)h(v),
\end{align}
where $\psi=\frac{w-w_0}{t}$. }
Since the particles are identical hard core point particles, at each collision they exchange their velocities. If, at each collision, one also exchanges the labels of the particles along with their velocities, one gets a gas of non-interacting point particles \cite{jepsen1965dynamics}. Since the quantity $\mathcal{N}_t$ is invariant under permutation of the labels of the hard-core point particles, one can use the non-interacting picture to compute the distribution $\mathcal{P}(n,t|x)$. Using such a picture,  it is easy to show that this distribution is given by the binomial distribution
\begin{align}
\mathcal{P}(n,t|x) &= \binom{N}{n} \left[\mathtt{p}_{r\ell}(x,t|0)\right]^n~\left[1-\mathtt{p}_{r\ell}(x,t|0)\right]^{N-n},\\ 
&= \binom{N}{n} \left(\frac{p_{r\ell}(x,t|0)}{N}\right)^n\left(1-\frac{p_{r\ell}(x,t|0)}{N}\right)^{N-n},
\end{align} 
{where $x=x_0+v_0t$ and $x_0=X_0(1-a\varrho_0)$.}
Using this probability, one can easily show that the probability of finding the quasiparticle at position $X$ at time $t$ starting from position $X_0<0$ is 
\begin{align}
P(X,t|X_0) &= \sum_{n=0}^N \delta_{n,\left \lfloor \frac{X-X_0-v_0t}{a} \right \rfloor}~\mathcal{P}(n,t|x), \label{def:P(Z,t|z_0)}\\
& =  \binom{N}{N(X,x_0)} \left(\frac{p_{r\ell}(x,t|0)}{N}\right)^{N(X,x_0)}  
\left(1-\frac{p_{r\ell}(x,t|0)}{N}\right)^{N-N(X,x_0)}, 
\end{align}
where
\begin{align}
N(X,x_0)=\frac{X-X_0-v_0t}{a}=\frac{X-x_0(1+a\varphi_0)-v_0t}{a},
\end{align}
and $\lfloor ~ \rfloor$ represents the floor function. Recall $X_0=x_0(1+a\varphi_0)$ [see Eq,~\eqref{mapp-z-Z}]. Now integrating $P(X,t|X_0)$ over $x_0$ with density $\varphi_0$ one gets the density 
profile of the special hard rods (those with velocity $v_0$)
\begin{align}
g(X,t) &= \int_{-L/2}^0dX_0~\varrho_0~P(X,t|X_0)=\int_{-L'/2}^0dx_0~\varphi_0~P(X,t|x_0(1+a\varphi_0)), \\ 
& = \varphi_0 \int_{-L'/2}^0dx_0~ 
 \left[
\binom{N}{N(X,x_0)} \left(\frac{p_{r\ell}(x_0+v_0t,t|0)}{N}\right)^{N(X,x_0)} \right. \nonumber \\
&~~~~~~~~~~~~~~~~~~~~~~~~~~\times~
\left. \left(1-\frac{p_{r\ell}(x_0+v_0t,t|0)}{N}\right)^{N-N(X,x_0)}
\right ]. 
\label{eq:g(Z,t)}
\end{align}

\begin{figure}[h]
    \centering
    \includegraphics[scale=0.65]{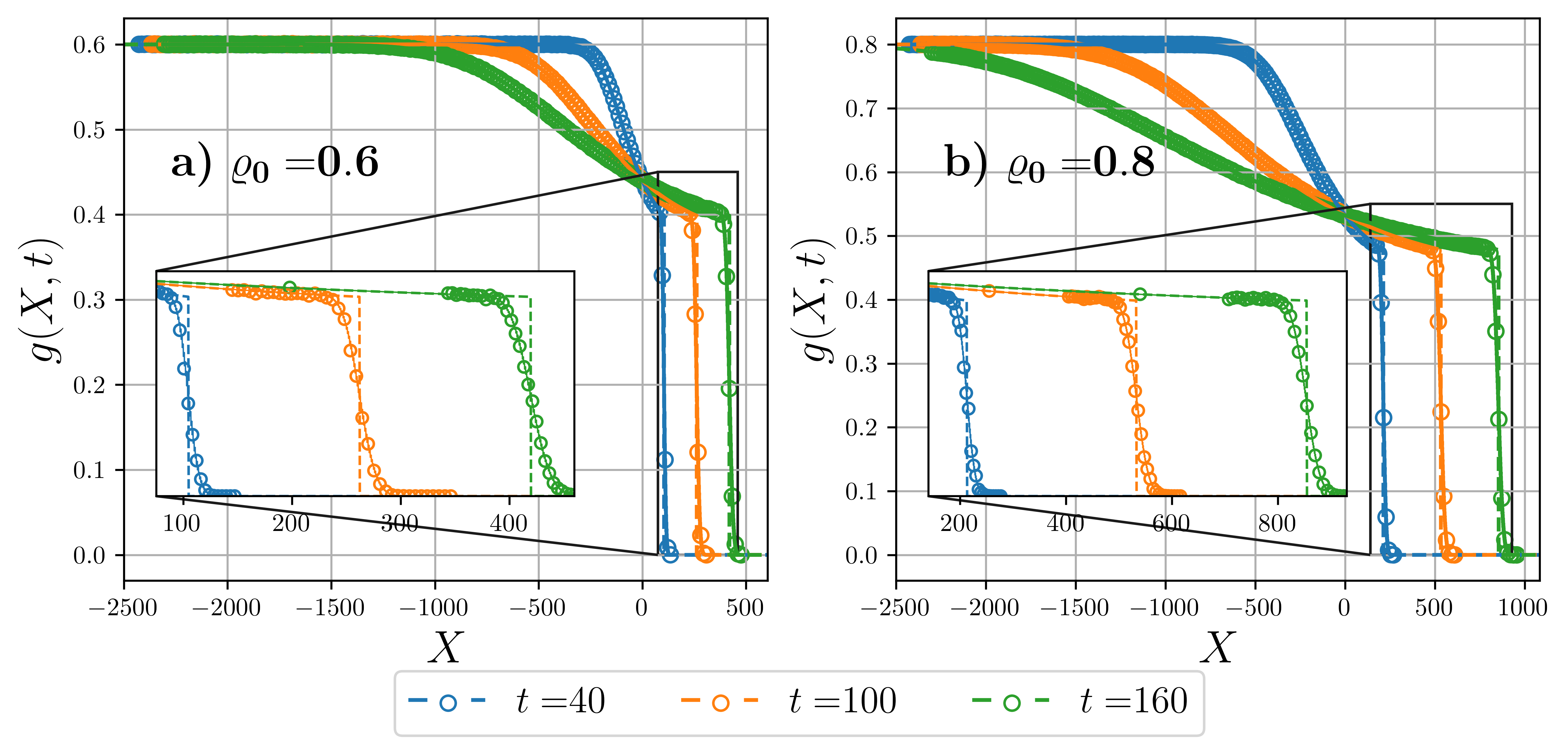}
    \caption{Plots  showing the density profiles for special hard rods at $t=40$ (blue), $t=100$ (orange) and $t=160$ (green) for $\varrho_0=0.6$ (left) and $\varrho_0=0.8$ (right). The circles represent simulation data. The solid lines correspond to the solution in Eq.~\eqref{eq:g(Z,t)} and the dashed lines correspond to the Euler solution in Eq.~\eqref{g(Z,t)-2}. The simulations are performed inside a 1D box of size $X\in[-2500,2500]$. Both microscopic as well as the Euler solution agree with the simulation data in both plots except at locations of the jumps. The density profiles are zoomed near the jump locations in the insets, where we  observe that the microscopic solution in Eq.~\eqref{eq:g(Z,t)} has an excellent agreement as compared to the Euler GHD solution. The times are chosen such that the effect from the collisions from the boundaries of the box are negligible and one can use theoretical expressions in Eq.~\eqref{eq:g(Z,t)} and Eq.~\eqref{g(Z,t)-2}. The simulation data are obtained after averaging over $10^5$ realizations.}
    \label{ICA-Special-HR-4k}
\end{figure}

\noindent
To take the thermodynamic limit $N \to \infty$, $L' \to \infty$ keeping $\varphi_0 =N/L'$ fixed, it seems convenient to perform this limit first on the MGF $\mathcal{Z}(\alpha,t|x_0)=\langle e^{-\iota \alpha X} |x_0\rangle$ and then perform the inverse Fourier transform.  The MGF simplifies as 
\begin{align}
\mathcal{Z}(\alpha,t|x_0) = \int dX~e^{-\iota \alpha X} P(X,t|X_0).
\end{align}
Inserting the form of $P(X,t|X_0)$ from Eq.~\eqref{def:P(Z,t|z_0)} and simplifying, we get 
\begin{align}
\mathcal{Z}(\alpha,t|x_0) = e^{-\iota \alpha (x_0(1+a\varphi_0)+v_0t)}~\left[ 1+\left(e^{-\iota \alpha a}-1\right)\frac{p_{r\ell}(x_0+v_0t,t|0)}{N}\right]^N,
\end{align}
which in the $N \to \infty$ limit provides,
\begin{align}
\mathcal{Z}(\alpha,t|x_0) =\exp \left[ -\iota \alpha(x_0(1+a\varphi_0)+v_0t) +\left(e^{-\iota \alpha a}-1\right)~p_{r\ell}(x_0+v_0t,t|0) \right]
\end{align}
Then $g(X,t)$ in the thermodynamic limit becomes
\begin{align}
\begin{split}
g(X,t) =\varphi_0 \int_{-\infty}^0dx_0 \int_{-\infty}^\infty \frac{d\alpha}{2\pi}~e^{\iota \alpha X}~
\exp &\Big[ -\iota \alpha (x_0(1+a\varphi_0)+v_0t) \\ 
&~~ +\left(e^{-\iota \alpha a}-1\right)~p_{r\ell}(x_0+v_0t,t|0) \Big].
\end{split}
\end{align}
{As done before, we approximate the factor $(e^{-\iota \alpha a}-1) \approx  -\iota \alpha a $ and take the $N\to \infty$ and $a \to 0$ limit keeping $Na$ fixed. We find the following} the Euler solution
\begin{align}
g(X,t) &=  \varphi_0 \int_{-\infty}^{v_0t}dx~\delta\left(X- \langle X \rangle(x)\right),~~\text{where}, \label{g(Z,t)-1}\\
\langle X\rangle(x) &=\left[ x+a(x-v_0t)\varphi_0+a p_{r\ell}(x,t|0) \right]. \label{<Z>(z)}
\end{align}
In the above we have used the variable transformation $x=x_0+v_0t$.
Performing the integral over $x$ yields 
\begin{align}
 g(X,t) &=\frac{\varphi_0 \Theta(v_0t-x^*(X))}{1+a\varphi_0+a[\partial_xp_{r\ell}(x,t|0)]_{x=x^*(X)}}, 
 \label{g(Z,t)-2}
\end{align}
where $x^*(X)$ is a solution of the equation 
\begin{align}
x+a\varphi_0(x-v_0t)+ap_{r\ell}(x,t|0)=X.
 \label{z*(Z)}
\end{align}
Note the expression of $g(X,t)$ in Eq.~\eqref{g(Z,t)-2} along with Eq.~\eqref{z*(Z)} is indeed the  Euler solution of the form in Eq.~\eqref{sol:euler-GHD-hr} with velocity $V$ integrated. One can also get this solution by directly solving the Euler GHD equation \eqref{eq:GHD} as done in \cite{singh2024thermalization}. 

In Fig.~\ref{ICA-Special-HR-4k}, we provide a numerical verification of the analytical result for the density $g(X,t)$ of the special hard rods for two values of the mass density $\varrho_0=0.6$ (left) and $\varrho_0=0.8$ (right). For comparison, we also plot the Euler solution from Eq.~\eqref{g(Z,t)-2} along with Eq.~\eqref{z*(Z)}. As noted in ~\cite{singh2024thermalization}, the Euler solution matches with the numerical solution everywhere except at the location of the jumps. The microscopic computation provides  exact solution for the density profile $g(X,t)$ in Eq.~\eqref{eq:g(Z,t)} which goes beyond the Euler solution and indeed agrees well with the numerical data as shown in the insets of both the plots in Fig.~\ref{ICA-Special-HR-4k} where we have {zoomed} in the regions near the jump (shock) locations at different times. \\

\subsection{Computation of the density profile $\rho_b(X,t)$ for background rods}
In this section we study the density profile of the background rods. Like in the case of special rods considered in the previous section, we look at the dynamics of a {background} quasiparticle. Such a rod with initial position, $X_0 > 0$ and velocity $v$, can undergo collisions with both the special rods approaching from the left, and the background rods approaching from either the left or the right. Any collision from the left (right) will make the quasiparticle jump by rod length $a$ to the left (right). Hence, at any time $t$ the position of the {background} quasiparticle having velocity, $v$ is given by:
\begin{equation}
    X(t) = X_0 + vt + a(N^{\rm (b)}_r - N^{\rm (b)}_\ell - N^{\rm (s)}) \ ,
\end{equation}
where $N_r^{\rm (b)}$ ($N_\ell^{\rm (b)}$) denotes the number of collisions with the background rods from the right (left) while $N^{\rm (s)}$ denotes the number of collisions with the special rods till time, $t$. The superscript `s' (`b') stands for special (background) hard rods while the subscript $r$ ($\ell$) stands for right (left). In the corresponding point particle picture, the same background rod is mapped to a background point particle with initial position, $x_0 = X_0 (1-a \varrho_0)$ moving with the same velocity, $v$ in a box of reduced length, $2L' = 2L(1-a\varrho_0) $. Till time $t$, a background {point particle} will encounter exactly the same number of particles as the number of collisions the corresponding {background} quasiparticle would undergo.  

To find the distribution for $X(t)$, we need to find the distribution of $\mathcal{P}(N^{\rm (b)}_r, N^{\rm (b)}_\ell, N^{\rm (s)}, t)$ which is easier to compute in the point particle picture. The collisions of the background point particles with the special point particles are independent of their collisions with the background point particles. Hence, we can write 
\begin{equation}
    \mathcal{P}(N^{\rm (b)}_r, N^{\rm (b)}_\ell, N^{\rm (s)}, t) = \mathcal{P}^{\rm (b)}(N^{\rm (b)}_r, N^{\rm (b)}_\ell, t) \times  \mathcal{P}^{\rm (s)}(N^{\rm (s)},t) \ ,
    \label{eq:mcalP(nr,nl,ns,t)}
\end{equation}
where $\mathcal{P}^{\rm (s)}(N^{\rm (s)},t)$ is the probability that the background point particle faces  exactly $N^{\rm (s)}$ encounter with special point particles till time, $t$ while $\mathcal{P}^{\rm (b)}(N_r^{\rm (b)},N_\ell^{\rm (b)},t)$ is the probability that the background point particle  encounters $N_r^{\rm (b)}$ and $N_\ell^{\rm (b)}$ background particles from the right and the left respectively till time $t$. The calculation of the probability $\mathcal{P}(N^{\rm (b)}_r= n_r, N^{\rm (b)}_\ell = n_\ell, N^{\rm (s)} =  n_s, t)$ is provided  in \ref{app:mcalP(nr,nl,ns,t)} in detail.

\begin{figure}[t]
    \centering
    \includegraphics[scale=0.6]{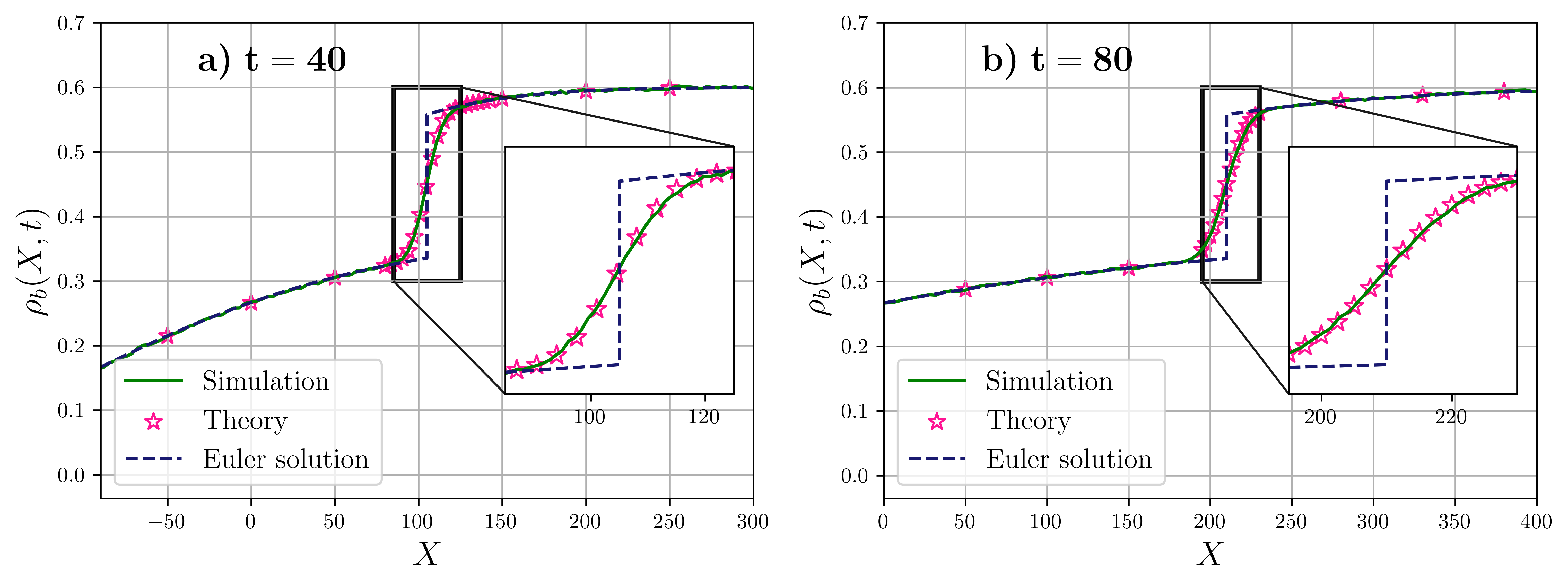}
    \caption{Plots showing a comparison of the density profile of background rods at $t =40$ and $t=80$ starting from the domain wall initial condition. The initial overall density of the  rods is $\varrho_0 = 0.6$. The velocity of the special rods are $v_0=1$ and the velocities of the background rods are chosen from Maxwell distribution in Eq.~\eqref{eq:max-dist} with $T=1.0$. The solid lines (green) are results from numerical simulation and the stars (deep pink) are obtained evaluating the theoretical result in Eq.~\eqref{eq-rhobackgrZtanalytic} numerically (see \ref{app:numer-eval} for details of numerical evaluation).  The simulation data are obtained after averaging over $10^5$ realizations. For comparison we also plot the Euler solution (dashed lines in midnight blue) of the GHD equation given in section (4.1) of ~\cite{singh2024thermalization}. We observe that both our microscopic solution as well as the Euler solution agree with numerical simulation data well except at the locations of the jumps. At this location, the density profiles, instead of having a discontinuity as predicted by the Euler solutions, is smooth and widened. In order to demonstrate the agreement between the microscopic solution and the Euler solution, we have zoomed the   the region near the jump locations in the insets. The simulations are performed inside a 1D box of size $X\in[-1250,1250]$ with $N = 750$ and $a = 1$.}
    \label{fig:Backgr}
\end{figure}

 Having found these probabilities explicitly in  Eq.~\eqref{appeq:mcalP(nr,nl,ns,t)}, one can  write the probability that a {background} quasiparticle with velocity $v$ moves to $X$ from $X_0$ at time $t$ as 
\begin{align}\label{summP}
    P^{\rm (b)}(X,t|X_0,v) = \sum_{n_s = 0}^{N} \ \sum_{n_r = 0}^{N - 1} \ \sum_{n_\ell = 0}^{N - n_r - 1} \delta_{  (n_r - n_\ell - n_s),\left\lfloor \frac{X - X_0 - vt}{a} \right \rfloor} \ \mathcal{P}^{\rm (s)} (n_s, t) \ \mathcal{P}^{\rm (b)} (n_r,n_\ell,t).   
\end{align}
Here one should note that the probabilities $\mathcal{P}^{\rm (s)}$ and $\mathcal{P}^{\rm (b)}$ also depend explicitly on $x=X_0(1-a\varrho_0)+vt$ and $v$ [see  Eq.~\eqref{appeq:mcalP(nr,nl,ns,t)}]. For simplicity of notations we have suppressed their explicit dependencies. 
\begin{figure}[t]
    \centering
    \includegraphics[scale = 0.7]{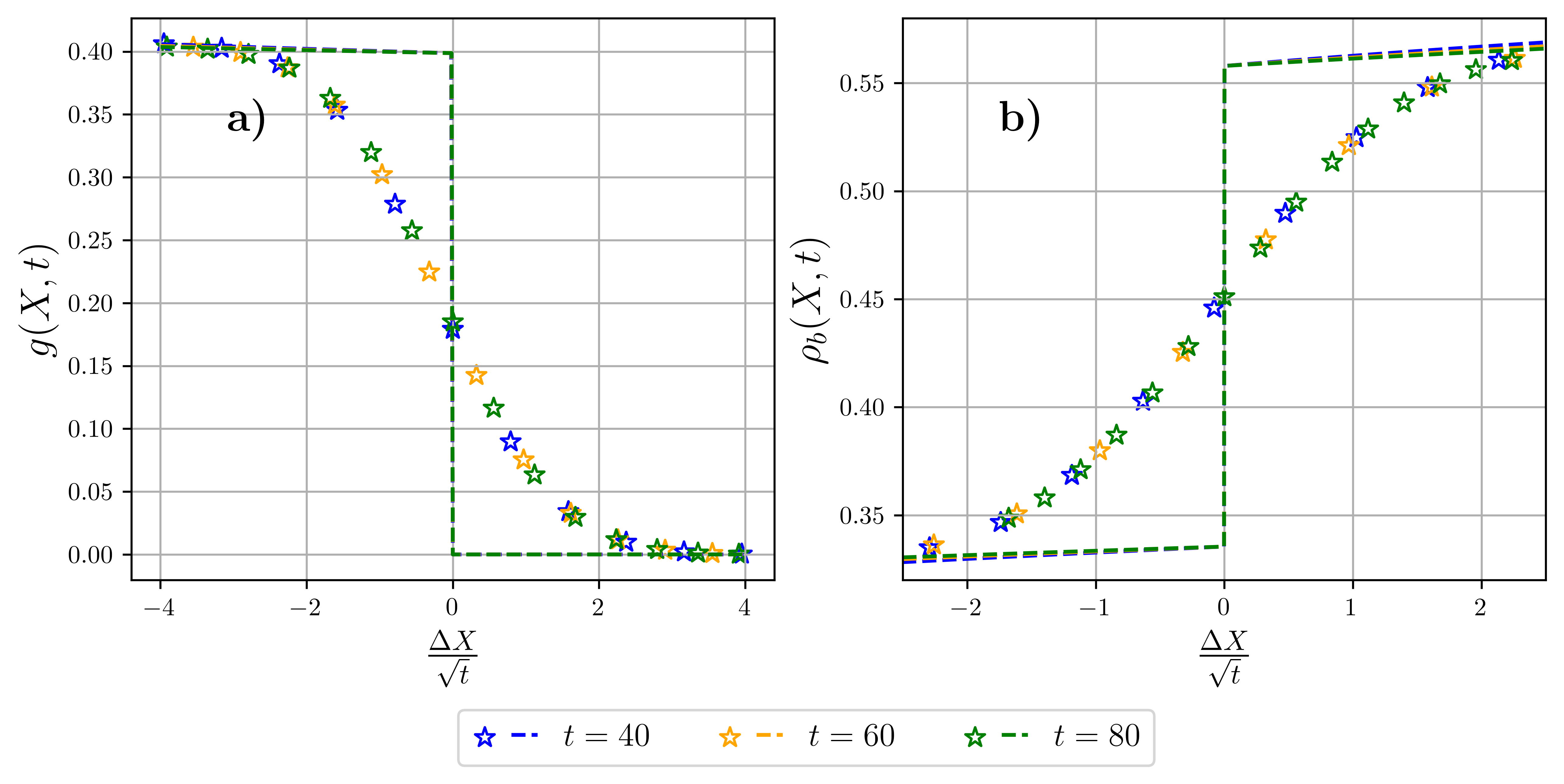}
    \caption{{Plot showing the scaling collapse of the mass density profiles $g(X,t)$ and $\rho_b (X,t)$, respectively, of the special and background rods (zoomed) around the location of the wall front $X_{\rm eu}(t)$ in the Euler solution. Here, $\Delta X = X - X_{\rm eu}$. The dashed lines represent the Euler solution and the stars represent the microscopic solution. This provides evidence that the microscopic solution captures the $\sqrt{t}$ spread of the density profile around the shock in the Euler solution. In plot (a) the parameters are $N = 1500$, $L = 2500$, $a = 1$, $\varrho_0 = 0.6$ and $X\in[-L,L]$. Whereas, in plot (b) the parameters are $N = 750$, $L = 1250$, $a = 1$, $\varrho_0 = 0.6$.}}
    \label{fig:scaling_collapse}
\end{figure}
The density, $\rho_b (X,t)$ of background rods at any time, $t$ is then obtained by integrating over the initial position $X_0$ and the velocity $v$
\begin{equation}\label{eq-rhobackgrZtanalytic}
    \rho_b (X,t) = \varrho_0 \int_{0}^\infty \text{d}X_0 \int_{-\infty}^\infty \text{d}v \ h(v) \ P^{\rm (b)}(X,t|X_0,v) \ ,
\end{equation}
where, $h(v)$ is the velocity distribution of the background rods. We numerically evaluate the background density $\rho_b (X,t)$ and compare with the same obtained from numerical simulation of the hard rod dynamics directly. The comparison between the two is given in Fig.~\ref{fig:Backgr} where we observe good agreement between the two.  As mentioned earlier,  the solution of the  Euler solution of the GHD equation for the background rods, starting from this domain wall initial configuration, was obtained recently in ~\cite{singh2024thermalization} (see section (4.1) of this paper) where it was shown that the initial jump at the interface with the special rods remains at all times and moves with an effective velocity. As shown in the inset of Fig.~\ref{fig:Backgr}, the numerically simulated density profile displays widening of this discontinuity which indeed is captured by our result in Eq.~\eqref{eq-rhobackgrZtanalytic} obtained microscopically. We also notice from Fig.~\ref{fig:Backgr} that the width of the shock increases with time and in \cite{singh2024thermalization} it was argued that the width grows as $\sqrt{t}$. {We demonstrate this fact numerically in Fig.~\ref{fig:scaling_collapse} where we plot the mass density profile $\rho_b(X,t)$ of the background rods as functions of  $\frac{(X-X_{\rm eu}(t))}{\sqrt{t}}$ at different times, where $X_{\rm eu}(t)$ is the displacement of the domain wall or the shock in the Euler solution from the initial position (which was at the origin). This displacement can be computed from by solving Eq.~\eqref{z*(Z)} for $X$ with $x=v_0t$. The excellent scaling collapse  in  Fig.~\ref{fig:scaling_collapse} for both special (left) and background (right) rods, verify the $\sim \sqrt{t}$ growth of the width of the domain wall front.}
We end this section by making the remark that, multiplying $v^\alpha$ in the integrand under the $v$ integral in Eq.~\eqref{eq-rhobackgrZtanalytic} one can also compute other conserved densities $q_\alpha(X,t)$ of the background rods.

{\color{black}
\section{ Numerical solution of the Navier-Stokes Eq.~\eqref{eq:GHD-NS} and comparison with the microscopic solution}
\label{sec:NS-solution}
In this section, we present numerical solution of the NS GHD equation \eqref{eq:GHD-NS} and compare it with the microscopic as well as the Euler solution for the mass density profile. It was argued in \cite{singh2024thermalization} that the effect of the NS term is better seen in the mass density profile when the initial phase space distribution has a component with singular (or highly narrow) velocity distribution. The domain wall initial condition given by Eq.~\eqref{ic:domain-wall} and studied in Sec.~\ref{sec:domain-wall} indeed has such a component. However, the discontinuity in the mass density profiles of both special and background rods in this initial condition is not suitable for numerical solution of the NS GHD equation \eqref{eq:GHD-NS}. For this reason, we consider a slightly smoother version of the domain wall initial condition. 
Let us consider $2N$ hard rods in a box of length $2L$. Eventually, we would consider $N \to \infty, L\to \infty$ limit. Now, we would consider the following initial condition (in the HR picture):
\begin{equation}
    f(X,V,t=0) = g^\mathcal{W}(X,t=0) \delta(V - V_0) \Theta(X+L) + f^\mathcal{W}_b (X,V,t=0) \Theta(L-X), \label{ic:domain-wall-smooth}
\end{equation}
where
\begin{subequations}
\label{eq:error-domain-wall}
    \begin{equation}
        g^\mathcal{W}(X,t=0) = \frac{\varrho_0}{2} \bigg(1 - {\rm Erf} \bigg[ \frac{X}{\mathcal{W}} \bigg] \bigg)
    \end{equation}
    \begin{equation}
        f^\mathcal{W}_b(X,V,t=0) = \frac{\varrho_0}{2} \bigg(1 + {\rm Erf} \bigg[ \frac{X}{\mathcal{W}} \bigg] \bigg) \ h(V) \ .
    \end{equation}
\end{subequations}
Note that in the limit $\mathcal{W} \to 0$, this initial condition indeed  reduces to the initial condition  in Eq.~\eqref{ic:domain-wall}. For non-zero $\mathcal{W}$, Eq.~\eqref{eq:error-domain-wall} is a somewhat smoother version of Eq.~\eqref{ic:domain-wall}. To solve the NS GHD equation \eqref{eq:GHD-NS}, we first separate the phase space density into a special component $g^\mathcal{W}(X,t)$ and a background component $ f^\mathcal{W}_b(X,V,t)$, resulting in a set of coupled partial differential equations (See Eq.~\eqref{eq:dom-GHD-NS}). We then solve these equations using finite difference method starting from the initial condition in Eq.~\eqref{ic:domain-wall-smooth}.  The details of the numerical procedure   are given in \ref{app:sol-dom_GHD-NS}. In Fig.~\ref{fig:comparison-error-init} we plot the mass density profiles of  the special rods, $g^\mathcal{W}(X,t)$ and of the background rods, $\rho_b^\mathcal{W}(X,t)$ obtained from the numerical solution of the NS GHD equation (orange dots). 
\begin{figure}[t]
    \centering
    \includegraphics[scale=0.6]{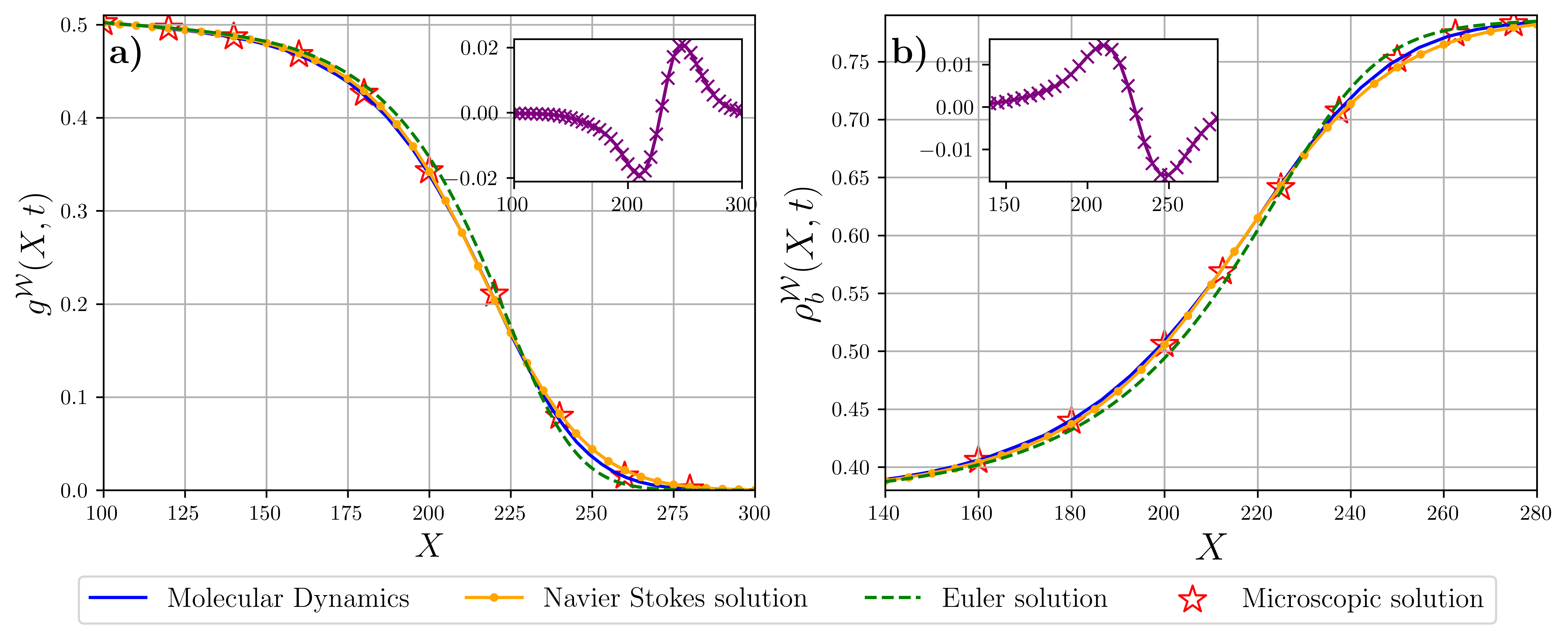}
    \caption{{Plot showing the comparison of the Navier-Stokes solution (orange dots) and the Euler solution (green dashed line) for the mass densities corresponding to (a) special hard rods and (b)background hard rods, at time, $t=40$ from the initial condition \eqref{eq:error-domain-wall}. We observe that the Navier-Stokes solution has a better agreement with molecular dynamics (blue solid line) than the Euler solution. The inset shows the deviation of the Navier-Stokes solution from the Euler solution. Moreover, we show that the microscopic solution (red stars) also agrees well with both Navier-Stokes solution and molecular dynamics. Here, $N=1000$, $L=1250$, $a=1$ and $X\in[-1250,1250]$.}}
    \label{fig:comparison-error-init}
\end{figure}
We want to compare this numerical solution with the Euler solution. For that, one should first write down the initial condition in PP coordinates as:
\begin{subequations}
\label{eq:error-domain-wall-PP}
    \begin{equation}
        g^{0,w}(x,t=0) = \frac{\varrho_0}{2 (1 - a \varrho_0)} \bigg(1 - {\rm Erf} \bigg[ \frac{x}{w} \bigg] \bigg)
    \end{equation}
    \begin{equation}
        f^{0,w}_b(x,V,t=0) = \frac{\varrho_0}{2 (1 - a \varrho_0)} \bigg(1 + {\rm Erf} \bigg[ \frac{x}{w} \bigg] \bigg) \ h(V) \ ,
    \end{equation}
\end{subequations}
where  $w = \mathcal{W} (1-a \varrho_0)$. Boosting these initial conditions with their respective velocities we get the following 
Euler solution in PP picture at later time, $t$:
\begin{subequations}
\label{eq:error-domain-wall-PP-euler}
    \begin{equation}
        g^{0,w}(x,t) = g^{0,w}(x-V_0 t,t=0) = \frac{\varrho_0}{2 (1 - a \varrho_0)} \bigg(1 - {\rm Erf} \bigg[ \frac{x - V_0 t}{w} \bigg] \bigg)
    \end{equation}
    \begin{equation}
        f^{0,w}_b(x,V,t) = f^{0,w}_b(x - Vt,V,t=0) = \frac{\varrho_0}{2 (1 - a \varrho_0)} \bigg(1 + {\rm Erf} \bigg[ \frac{x-Vt}{w} \bigg] \bigg) \ h(V) \ .
    \end{equation}
\end{subequations}
Now using the mapping in Eqs.~(\ref{sol:euler-micro}-\ref{hd-transfrmtn}), one obtains the solution of the Euler solution for hard rods. 
In Fig.~\ref{fig:comparison-error-init}, we observe that the Euler solution does not agree with the Navier-Stokes solution near the domain wall front. The Navier-Stokes solution for both the components shows a larger spread of the wall front than the Euler solution. This is also demonstrated further in the insets where we plot the difference between the two. 

Following the process outlined in the previous section, one can also compute the microscopic solution.
In Fig.~\ref{fig:comparison-error-init}, we also show a comparison between the microscopic solution (red stars) of the mass densities of both the special and the background components with the molecular dynamics simulations (blue solid line), the Euler solution (green dashed line) and the Navier-Stokes solution (orange dots). Interestingly, we see that in addition to agreeing well with results from molecular dynamics as was demonstrated on several contexts in the previous sections, the microscopic solution agrees well with Navier-Stokes solution as well.
This further demonstrates how the solution of the GHD equation with NS correction can approximately describe the microscopic solution and display features at length scales smaller than the Euler scale. 

\section{Conclusion}
\label{sec:conclusion}
The macroscopic evolution of the densities of many interacting particle systems from non-equilibrium initial state and its approach to equilibrium is an important problem. Such a problem is often studied by hydrodynamics.  While in the context of non-integrable systems such studies are quite old, but in the context of integrable systems, they have grown a large interest  in last decade. In this paper we have studied the macroscopic evolution of non-equilibrium initial state in a collection of {hard rod}s in one dimension, but from a microscopic point of view. Obtaining exact and explicit solutions of the conserved density profiles for finite-sized systems, we have demonstrated how in the thermodynamic limit one can get the Euler solutions of the GHD equations. Furthermore, we showed how one can get corrections to the Euler solutions and thus go beyond Euler solutions. We showed that these corrections are different for quenched and annealed initial conditions. We have demonstrated this difference quantitatively in the specific problem of tracer diffusion in the inhomogeneous background. Finally, we applied our microscopic approach to the case of a specially designed domain wall initial condition, in which there are locations of the jumps in the Euler solutions. We demonstrate that our solution can  describe the density profiles accurately even at the locations of the jumps (shocks) where the Euler solution fails. 

Our microscopic approach can be explored further in different directions. The non-equilibrium space-time correlation functions of the conserved densities provides information about the fluctuations as well as the relaxation of the system to the stationary state. Recently, for integrable systems to compute such correlations, a ballistic macroscopic fluctuation theory has been developed, which makes precise computable predictions \cite{doyon2023emergence,doyon2023ballistic}. It would be interesting to reproduce those predictions from a microscopic approach for {hard rod} systems. It is known that the Euler solutions do not produce any entropy. It would also be interesting to see how the microscopic solutions on macroscopic scale incur entropy with time. Recently, for integrable systems another  question has drawn a lot of interest: if a few of the conserved laws are broken (while still having extensive number of them), then how would the dynamics at macroscale get modified? It would be interesting to explore this question in the context of {hard rod} system.

\section{Acknowledgements}
We would like to thank Samriddhi Sankar Ray for useful discussions and help on the numerical solution of the GHD equation with NS terms. AK would like to acknowledge the support of DST, Government of India Grant under Project No. CRG/2021/002455 and the MATRICS grant MTR/2021/000350 from the SERB, DST, Government of India. MJP and AK acknowledge the Department of Atomic Energy, Government of India, for their support under Project No. RTI4001.

\appendix
\section{Derivation of Eq.~\eqref{def:q_p-q-m}}
\label{derivation of mttP(z,v,t)-line}
We start with the probability 
\begin{align}
\mathscr{P}_i(z,v,t|{\bf x},{\bf u})  
& =\int_{-\infty}^zdy_{i-1} ... \int_{-\infty}^{y_2}dy_1 ~\times\int_z^\infty dy_{i+1}{...\int_{y_{N-1}}^\infty dy_N}\left[\prod_{k=1}^N\int_{-\infty}^\infty dv_k\right] \delta(v_i-v) \notag \\ 
&~~~~~~\times~~~\mathbb{G}_N(y_1,...,y_{i-1},z,y_{i+1},...,y_N;{\bf v};t|{\bf x},{\bf u},0), \notag \\
&=\int_{-\infty}^zdy_{i-1} ... \int_{-\infty}^{y_2}dy_1 ~\times~\int_z^\infty dy_{i+1}{...\int_{y_{N-1}}^\infty dy_N}\notag \\ 
&~~~~~~\times~~\sum_{\tau \in \sigma_N} g(z,v,t|x_{\tau(i)},u_{\tau(i)},0),  \prod_{\substack{k=1\\k\ne i}}^N \int_{-\infty}^\infty dv_k g(y_k,v_k,t|x_{\tau(k)},u_{\tau(k)},0),
\end{align}
where $\sigma_N$ represents the set of permutations over $N$ distinct elements.
The summation over these permutations can be performed into three steps. In the first step, we sum over {sets} $S_\ell=[\tau(1),...,\tau(i-1)]$, ~$\tau(i)$ and $S_r=[\tau(i+1),...,\tau(N)]$ and in the second step we sum over permutations $\tau_\ell \in S_\ell$ and $\tau_r \in S_r$ respectively. 
\begin{align}
\mathscr{P}_i(z,v,t|{\bf x},&{\bf u})  
=\sum_{S_\ell, m, S_r} g(z,v,t|x_m,u_m,0)  \notag  \\ 
&~~\times~ \sum_{\tau_\ell \in \sigma_{(i-1)}} \int_{-\infty}^zdy_{i-1} ... \int_{-\infty}^{y_2}dy_1 \prod_{k=1}^{i-1} \int_{-\infty}^\infty dv_k g(y_k,v_k,t|x_{S_\ell[\tau_\ell(k)]},u_{S_\ell[\tau_\ell(k)]},0) \notag  \\ 
&~~\times~ \sum_{\tau_r \in \sigma_{(N-i)}}  ...\int_z^{{\infty}} dy_{i+1}{...\int_{y_{N-1}}^\infty dy_N} \prod_{j=i+1}^N g(y_k,v_k,t|x_{S_r[\tau_r(j)]},u_{S_r[\tau_r(j)]},0), \label{derv:mttP-1} \\ 
&~~=\sum_{S_\ell, m, S_r}  g(z,v,t|x_m,u_m,0)~\notag \\ 
&~~~~~~~~~~~~~\times~\prod_{k=1}^{i-1} g_{<}(z,t|x_{S_\ell[k]},u_{S_\ell[k]},0) 
\prod_{j=i+1}^{N} g_{>}(z,t|x_{S_r[j]},u_{S_r[j]},0), \notag
\end{align}
where $x_{S_{\ell}[k]}$ represents the $k^{\rm th}$ element of the set of $(i-1)$ labels in the partition $S_\ell$. Similar meaning holds for  $x_{S_{r}[k]}$.
In the  fourth line of Eq.~\eqref{derv:mttP-1} we have used the following results
\begin{align}
\begin{split}
\sum_{\tau_\ell \in \sigma_{(i-1)}} \int_{-\infty}^zdy_{i-1} ... \int_{-\infty}^{y_2}dy_1 \prod_{k=1}^{i-1} \int_{-\infty}^\infty & dv_k g(y_k,v_k,t|x_{S_\ell[\tau_\ell(k)]},u_{S_\ell[\tau_\ell(k)]},0) \nonumber \\ 
&~~~~
= \prod_{k=1}^{i-1} g_{<}(z,t|x_{S_\ell[k]},u_{S_\ell[k]},0), \\
\sum_{\tau_r \in \sigma_{(N-i)}} \int_z^{{\infty}} dy_{i+1}{...\int_{y_{N-1}}^\infty dy_N}\prod_{k=i+1}^N &g(y_k,v_k,t|x_{S_r[\tau_r(k)]},u_{S_r[\tau_r(k)]},0) 
\nonumber \\ 
&~~~~= \prod_{k=i+1}^{N} 
g_{>}(z,t|x_{S_r[k]},u_{S_r[k]},0), 
\end{split}
\notag
\end{align}
where 
\begin{align}
\begin{split}
& g_{<}(z,t|x,u,0) = \int_{-\infty}^z dy\int_{-\infty}^\infty dv~g(y,v,t|x,u,0), \\ 
& \text{and}~~g_{>}(z,t|x,u,0) =1-g_{<}(z,t|x,u,0).
\end{split}
\label{def:q_p-q-m-ap} 
\end{align}
\subsection{Annealed Case:}
\label{ap:annealed-case}
Now to obtain $\mathtt{P}_i(z,v,t)$ defined in Eq.~\eqref{mttP(z,v,t)} we need to average over the initial conditions in $\mathscr{P}_i(z,v,t|{\bf x},{\bf u})  $
\begin{align}
\mathtt{P}_i(z,v,t) &=\prod_{k=1}^N  \int dx_k \int du_k~\mathscr{P}_i(z,v,t|{\bf x},{\bf u})  ~ \mathbb{P}_{a}(\{x_{j},v_{j}\},0), \notag \\ 
&=\prod_{k=1}^N  \int dx_k \int du_k~\mathscr{P}_i(z,v,t|{\bf x},{\bf u})  ~ \prod_{j=1}^N \mu_{a}(x_j,u_j)
\end{align}
 {Note in the last line we have used the fact that the variables $\{ x_i,~u_i\}$  are independent and identically distributed.}
Inserting the expression of $\mathscr{P}_i(z,v,t|{\bf x},{\bf u})  $ from Eq.~\eqref{derv:mttP-1} we get 
\begin{align}
\mathtt{P}_i(z,v,t) &=\sum_{S_\ell, m, S_r}  g^0(z,v,t)~\prod_{k=1}^{i-1} q(z,t) 
\prod_{k=i+1}^{N} [1-q(z,t)],~~~\text{where} \label{ex:mttP(z,v,t)}\\
g^0(z,v,t) &= \int dx \int du~g(z,v,t|x,u,0)~\mu_{a}(x,u), \label{def:g^0-g-ap} \\ 
q(z,t) &=  \int dx \int du~g_{<}(z,t|x,u,0)~\mu_{a}(x,u). \label{def:q-gm-ap}
\end{align}
Now it is easy to simplify the expression of $\mathtt{P}_i(z,v,t)$ in Eq.~\eqref{ex:mttP(z,v,t)} and one finds
\begin{align}
\mathtt{P}_i(z,v,t) &= g^0(z,v,t)~ \left[ q(z,t) \right]^{i-1}
\left[(1-q(z,t)\right]^{N-i}~\left[ \sum_{S_\ell, m, S_r}1 \right], \notag \\ 
&= g^0(z,v,t)~ \left[ q(z,t) \right]^{i-1}
\left[(1-q(z,t)\right]^{N-i}~\left[ N \binom{N-1}{i-1}\right].
\end{align}

\section{Computation of the joint probability $\mathcal{P}(N^{(b)}_r, N^{(b)}_\ell, N^{(s)}, t)$ in Eq.~\eqref{eq:mcalP(nr,nl,ns,t)}}
\label{app:mcalP(nr,nl,ns,t)}
To compute these probabilities, we first note that $N^{\rm (s)}$ is also equal to the number of special point particles that  reach positions above $z=z_0+vt$ at time $t$. Similarly, $N^{\rm (b)}_{r}$ ($N^{\rm (b)}_{\ell}$) also denotes the number of background point particles that starting from positions above (below) $x_0=X_0(1-a\varrho_0)$ reach positions below (above) $x=x_0+vt$ at time $t$. Let $\mathtt{p}_{\ell r}^{\rm (s)}$ represent the probability for a single special point particle to go from positions $x\le0$ to $x\ge x_0+vt$. It is given by 
\begin{align}
  \mathtt{p}^{\rm (s)}_{\ell r} = \frac{\varphi_0}{N}(v_0 t - x_0 - vt) ~\Theta(v_0 t - x_0 - v t). 
  \label{p^(s)_lr}
\end{align}
On the other hand, let $\mathtt{p}^{\rm (b)}_{\ell r}$ ($\mathtt{p}^{\rm (b)}_{r\ell}$) represents the probability for a single background point particle to go from positions below (above) $x_0$ to positions above (below) $x=x_0+vt$ at time $t$. From the propagator of a single background point particle $\mathfrak{g}(y,t|\bar{x},0)=\frac{1}{t}h\left(\frac{y-\bar{x}}{t}\right)$ [see Eq.~\eqref{propagator}], these probabilities can be computed easily and they are given by 
\begin{align}
    \mathtt{p}^{\rm (b)}_{\ell r}(x,v,t)&=\frac{t~\varphi_0}{N}~\int_v^{x/t}du~\int_u^\infty du'~ h(u'), \label{p^(b)_lr} \\
    \mathtt{p}^{\rm (b)}_{r \ell}(x,v,t)&=\frac{t~\varphi_0}{N}~\int_v^{\infty}du~\int_{-\infty}^{u} du'~ h(u'),\label{p^(b)_rl}
\end{align}
where, recall $\varphi_0=\frac{\varrho_0}{1-a\varrho_0}$.
In terms of these probabilities, one can write the probabilities $\mathcal{P}^{(s)}(N^{\rm (s)},t)$ and $\mathcal{P}^{\rm (b)}(N_r^{\rm (b)},N_\ell^{\rm (b)},t)$ as the following multinomial distributions
\begin{align}
\begin{split}
    \mathcal{P}^{\rm (s)}( n_s,t) &= \binom{N}{n_s} \left( \mathtt{p}^{\rm (s)}_{\ell r} (x,v,t) \right)^{n_s} \left(1- \mathtt{p}^{\rm (s)}_{\ell r} (x,v,t) \right)^{N - n_s} \ , \\
    \mathcal{P}^{\rm (b)}(n_r, n_\ell,&t) = \frac{(N-1)!}{n_\ell! \ n_r! \ (N - n_\ell - n_r - 1)!} \left( \mathtt{p}^{(b)}_{r \ell} (x,v,t) \right)^{n_r} \left( \mathtt{p}^{(b)}_{\ell r} (x,v,t) \right)^{n_\ell} \\
&~~~~~~~~~~~~~~~~~~~~\times ~~
\left(1 - \mathtt{p}^{(b)}_{\ell r} (x,v,t) - \mathtt{p}^{(b)}_{r \ell} (x,v,t) \right)^{N - n_r - n_\ell - 1} \ ,
\end{split}, \label{appeq:mcalP(nr,nl,ns,t)} 
\end{align} 
where $x=X_0(1-a\varrho_0)+vt$. Once again to write the above expressions we have followed the procedure in Sec.~\ref{sec:tracer-anld-ic}.
\section{Numerical evaluation of $\rho_b(Z,t)$ in Eq.~\eqref{eq-rhobackgrZtanalytic}} \label{app:numer-eval}
In this appendix we provide details about the numerical evaluation of the integrals in Eq.~\eqref{eq-rhobackgrZtanalytic}. It seems to be more convenient to perform the integral over $X_0$ and $v$ first in this and then perform the multiple sums in the expression of $P^{\rm (b)}(X,t|X_0,v)$ in Eq.~\eqref{summP}. After performing the integration over $X_0$, the integrand for integral over $v$ looks like the following (without the summation):
\begin{equation}\label{bbPmat}
    \bar{\mathbb{P}}(v,n_s,n_r,n_\ell) = \varrho_0  \ h(v) \ \mathcal{P}^{\rm (s)} (n_s, t) \ \mathcal{P}^{\rm (b)} (n_r,n_\ell,t) \biggr\vert_{X_0 = X - vt - (n_r - n_\ell - n_s)a} \ ,
\end{equation}
which is a function of $v$ for every triplet $(n_s,n_r,n_\ell)$. Here, $h(v)$ is a Gaussian function in $v$ and hence it decays very fast with increasing  $\vert v \vert$. This allows us to work with a smaller range of integration for $v$. The integration over $v$ can then be performed with Simpson's rule or any other suitable integration scheme. To reduce time consumption, we can perform the integration over $v$ for each triplet $(n_s,n_r,n_\ell)$ parallelly. This is achieved by noting that for each value of $v$, Eq. \eqref{bbPmat} denotes an element of a 3D matrix of shape $N \times (N-1) \times (N -1)$ indexed by $n_s, n_r \text{ and } n_\ell$ respectively. It is hence possible to define a function that takes an input $v'$ for the value of $v$ and returns the 3D array $\mathbb{P}$ evaluated at $v = v'$. Using this function, one can apply the Simpson's rule of integration for each triplet $(n_s,n_r,n_\ell)$ parallelly. At the end, we get the result of integration over $v$ for all the triplets in a matrix form. The multiple summation then reduces to the sum of the elements located on and below (or above) one of the diagonals of the resultant 3D matrix after integration.  

{
\section{Numerical evaluation of densities ($\rho^{\mathcal{W}}_{b}$ and $g^{\mathcal{W}}$) from GHD NS equation} \label{app:sol-dom_GHD-NS}
Similar to what was discussed in Sec.~\ref{sec:domain-wall}, there will be two parts in the phase space distribution function, $f(X,V,t)$,
\begin{equation}
    f(X,V,t) = g(X,t) \delta(V-V_0) + f_b (X,V,t) \ .
\end{equation}
When we substitute this form of $f(X,V,t)$ into the GHD equation with NS corrections, Eq.~\eqref{eq:GHD-NS}, and separate the special and the background components, we get
\begin{subequations}
\label{eq:dom-GHD-NS}
    \begin{equation}
    \begin{split}
        &\partial_t g (X,t) + \partial_X V_{\rm eff} (X,V_0,t) g(X,t) \\ &= \frac{a^2}{2(1-a \rho (X,t))} \int \text{d}W \vert V_0 - W \vert \Bigg[\bigg\{ f_b (X,W,t) \partial^2_X g(X,t) - g(X,t) \partial^2_X f_b (X,W,t) \bigg\}  \\[3pt]
        & \ \ \ \ \ \ \ \ \ \ \ \ \ \ + \frac{a \ \partial_X \rho(X,t)}{1 - a \rho(X,t)} \bigg\{f_b (X,W,t) \partial_X g(X,t) - g(X,t) \partial_X f_b (X,W,t)   \bigg\}  \Bigg] \ ,
    \end{split}
    \end{equation}
    \begin{equation}
        \begin{split}
            &\partial_t f_b(X,V,t) + \partial_X V_{\rm eff} (X,V,t) f_b (X,V,t) \\ &= \partial_X \Bigg[ \frac{a^2}{2(1-a\rho(X,t))} \int \text{d}W \vert V-W \vert \bigg\{f_b(X,W,t) \partial_X f_b (X,V,t) - f_b(X,V,t) \partial_X f_b(X,W,t)  \bigg\} \\
            & \ \ \ \ \ \ \ \ \ \ \ \ \ \ \ \ + \frac{a^2}{2(1 - a \rho(X,t))} \vert V-V_0 \vert \bigg\{ g(X,t) \ \partial_X f_b (X,V,t) - f_b (X,V,t) \partial_X g(X,t)  \bigg\}  \Bigg] \ .
        \end{split}
    \end{equation}
\end{subequations}
So, we obtain a set of coupled partial differential equations. We solve these for the initial conditions given in Eq.~\eqref{ic:domain-wall-smooth} along with Eq.~\eqref{eq:error-domain-wall}. We note that in the limit $\mathcal{W} \to 0$, we recover domain wall initial condition. To facilitate easy comparison with the domain wall initial condition, we choose to work with the same parameters ($a,L,N,\rho_0$) as we used for the latter. However, we scale the length and time by 1000 so that we can work with a smaller grid (hence facilitating less memory consumption), keeping the same scale for the velocity as before. For example, the parameters $L=1250,a = 1, N = 2000, \rho_0 = 0.8, t = 40, v_0 = 1, T = 1 $ for the domain wall initial condition are scaled to $L=1.25, a = 0.001, N = 2000, \rho_0 = 800, t = 0.04, v_0 = 1, T = 1$ for the new initial condition. \\[3pt] 
We work with a grid in position space for $g(X,t)$ with grid spacing, $\text{d}X = 0.005$ and a grid in phase space ($X,V$) for $f_b (X,V,t)$ with grid spacing $\text{d}X = 0.005, \text{d}V = 0.01$. We use fourth order Runge-Kutta scheme for time evolution with $\text{d}t = 5 \times 10^{-6}$, and we use second order accurate finite difference scheme to evaluate the $X$ derivatives. For integration in $V$, we use the standard rectangular integration scheme.
}
\begin{center}
\line(1,0){250}
\end{center}

\section*{References}

\bibliographystyle{unsrt}
\bibliography{references}

\end{document}